\documentclass[preprint,usenatbib,twocolumn,linenumbers]{aastex63}

\usepackage{newtxtext,newtxmath}
\usepackage{hyperref}
\usepackage[T1]{fontenc}
\usepackage{ae,aecompl}
\usepackage{graphicx}	% Including figure files
\usepackage{amssymb}	% Extra maths symbols
\usepackage[utf8]{inputenc}
\usepackage{mathtools}
\usepackage{color}      
\usepackage[caption=false]{subfig}
\usepackage{threeparttable}
\usepackage{lipsum}
\usepackage{float}
\usepackage{blindtext}
\usepackage[normalem]{ulem}
\usepackage{subfiles}
\nolinenumbers
\newcommand{\mi}[1]{\textsf{m12i}}
\newcommand{\mf}[1]{\textsf{m12f}}
\newcommand{\mw}[1]{\textsf{m12w}}

\newcommand{\ucdavis}{Department of Physics \& Astronomy, University of California, Davis, CA 95616, USA}
\newcommand{\upenn}{Department of Physics \& Astronomy, University of Pennsylvania, 209 S 33rd St, Philadelphia, PA 19104, USA}
\newcommand{\cca}{Center for Computational Astrophysics, Flatiron Institute, 162 5th Ave, New York, NY 10010, USA}

\received{2022 July 27}
\revised{2022 September 13}
\accepted{2022 September 20}
\accepted{APJ}
\shortauthors{Arora et al.}

\begin{document}
\title{On the stability of tidal streams in action space}

% The list of authors, and the short list which is used in the headers.
% If you need two or more lines of authors, add an extra line using \newauthor
\correspondingauthor{Arpit Arora}
\email{arora125@sas.upenn.edu}

\author[0000-0002-8354-7356]{Arpit Arora}
\affiliation{\upenn}

\author[0000-0003-3939-3297]{Robyn E. Sanderson}
\affiliation{\cca}
\affiliation{\upenn}

\author[0000-0001-5214-8822]{Nondh Panithanpaisal}
\affiliation{\upenn}

\author[0000-0002-6993-0826]{Emily C. Cunningham}
\affiliation{\cca}

\author[0000-0003-0603-8942]{Andrew Wetzel}
\affiliation{\ucdavis}

\author[0000-0001-7107-1744]{Nicol\'as Garavito-Camargo}
\affiliation{\cca}

\begin{abstract}
\nolinenumbers
In the Gaia era it is increasingly apparent that traditional static, parameterized models are insufficient to describe the mass distribution of our complex, dynamically evolving Milky Way (MW). In this work, we compare different time-evolving and time-independent representations of the gravitational potentials of simulated MW-mass galaxies from the FIRE-2 suite of cosmological-baryonic simulations. Using these potentials, we calculate actions for star particles in tidal streams around three galaxies with varying merger histories at each snapshot from 7 Gyr ago to the present day. We determine the action-space coherence preserved by each model using the Kullback-Leibler divergence to gauge the degree of clustering in actions and the relative stability of the clusters over time. We find that all models produce a clustered action space for simulations with no significant mergers. However, a massive (mass ratio prior to infall more similar than 1:8) interacting galaxy not present in the model will result in mischaracterized orbits for stars most affected by the interaction. The locations of the action space clusters (i.e. the orbits of the stream stars) are only preserved by the time-evolving model, while the time-independent models can lose significant amounts of information as soon as 0.5--1 Gyr ago, even if the system does not undergo a significant merger. Our results imply that reverse-integration of stream orbits in the MW using a fixed potential is likely to give incorrect results if integrated longer than 0.5 Gyr into the past.
\end{abstract}


\section{Introduction}

In the modern era of surveys like Gaia \citep{gaia2016}, Sloan Digital Sky Survey \citep{majewski2017apache, SDSSV2017}, and Rubin-LSST \citep{ivezic2019lsst} with high Galactic resolution, it is becoming obvious that our current static parametric models are unable to accurately describe the dynamically complex and evolving Milky Way (MW) halo. For years, researchers have used analytic density profiles, specified in terms of a handful of intuitive parameters, in order to understand the mass and shape of the MW \citep[e.g.,][]{McMillan_anly,bland2016galaxy,sand_anly}; a crucial step in constraining the nature of dark matter (DM) \citep[e.g.,][]{mao2015,robles2017sidm,thomas2018,fitts2019dwarf, sameie2021central} and understanding the formation of the MW \citep[e.g.,][]{Wechsler_2002,ludlow2013mass,bonaca2020timing,naidu2020evidence,LiTing2022}. These parametric models are efficient in approximating the present-day mass distribution of galaxies that match their symmetry and scaling assumptions, but they lack the necessary flexibility required to model galaxies with complicated {ongoing} mergers---and it is clear that the MW falls under this description, with a massive early merger (Gaia-Sausage-Enceladus \citep{helmi2018merger,belokurov2018co}) followed by a quiescent period and the current merger with massive infalling satellites such as the Large Magellanic Cloud (LMC) \citep{besla2007magellanic,kallivayalil2013third,penarrubia2015_LMC,vasiliev2021tango} and the progenitor of the Sagittarius stream (Sgr dSph) \citep{ibata1994,newberg2002ghost,kirby2013_Sag} {that} change the potential of the MW in nonadiabatic ways \citep[e.g.,][]{antoja2011understanding,laporte2018influence,erkal2019total,petersen2020reflex,garavito2021LMCimpact,vasiliev2021tango,lilleengen2022effect}. The time-independent, relatively symmetric parametric models previously used to describe the MW's potential fail to capture this evolution. {Recently, \citet{d2022uncertainties} used rigid spherical potentials for the MW-mass host and the most recent accreted satellite to backward integrate orbits of subhalos from their present-day positions. The found large uncertainties within the recovered orbits with respect to the true orbits.} In order for a model to accurately describe the global potentials, it should be able to incorporate perturbations caused by mergers in a Galactic halo. At the same time, the amount of information available to constrain potential models is far larger, and spans a larger volume, than ever before. Together these developments argue for establishing a new framework to model the global MW potential profile that is not limited by parameterization of quantities such as radius and mass, but takes into account the time evolution of halos.  One alternative and improved method is the multipole potential model based on basis function expansions (BFE), which can specify potentials of an arbitrary density distribution to high accuracy \citep{Hernquist1992, lowing_BFE} and can be used to describe MW-like halos. Given the MW's dynamic nature, we expect that a time-evolving potential model constructed using BFE will perform better than traditional parametric static potential models. Much work has already been done on the feasibility of orbit reconstruction in a time-dependent potential based on BFE. \citet{ngan2015simulating} simulated orbits after tidal disruption of a star cluster using BFE under a smooth halo and a lumpy halo, they found that two orbits were similar in orbital positions and had similar apo- and pericentric distances. Later, \citet{sanders2020models} tested a true time-dependent BFE-based model and quantified the error in orbit reconstruction and showed high fidelity orbit reconstruction is attainable using BFE. Previous work focused on reconstructing and comparing the phase-space information of tidally disrupted star clusters of dwarf galaxies, but a lot of modern techniques use actions as proxies for preserving orbits \citep[e.g.,][]{sanderson2015action,sanderson2017modeling,o2020velocity,Stella2020,trick2021identifying,}. We are specifically interested in how well different models preserve the actions of stellar streams and the orbits of halo objects in realistic simulations of MW-like galaxies. 

Stellar streams are formed when stars in a globular cluster or a dwarf galaxy are accreted into a more massive host galaxy due to tidal force from the gravitational potential of the host \citep{helmi1999building}. These streams were first predicted from a galaxy merger simulation by \citep{toomre1972galactic} followed by the first observational discovery by \citep{ibata1994, johnston1995disruption, newberg2002ghost} of a stream from the ongoing Sgr dSph merger, streams in globular clusters, and the Helmi streams \citep{Grillmair1995,helmi1999building}. These streams are excellent probes of the shape of the MW's radial mass profile \cite[e.g.,][]{koposov2010constraining, bonaca2014milky, pearson2015,sanderson2017modeling, Stella2020}, its symmetry, \cite[e.g.,][]{law2010sagittarius,Malhan_2019} and, the orientation of its symmetry axes \citep{vera2013constraints, vasiliev2021tango, Panithanpaisal_DMorient}.

When a model is a good representation of the true potential it can be used to transform the stream's phase-space coordinates to the action-angle variables. These streams are highly clustered in action space \citep{helmi1999building, sanderson2015action} because they arise from the tidal disruption of globular clusters or dwarf galaxies, which have much smaller initial phase-space volumes than the MW. 

Mass models of the MW are also used to explore the past and future orbits of satellite galaxies, which could affect their formation histories \citep{wetzel2015satellite,tomozeiu2016evolution,buck2019nihao} and our conclusions about their masses pre-infall \citep{mayer2001metamorphosis, lokas2010inner} and infall times \citep{sohn2020hst}. The most famous case is that of the Large Magellanic Cloud \citep{besla2007magellanic,kallivayalil2013third}, but connections between orbits and evolution have been explored for various other dwarfs as well, including Sgr \citep{johnston1995disruption,ibata1996kinematics,johnston1999constraining,jiang2000orbit} and Antlia II \citep{chakrabarti2019antlia}. The development of high-precision astrometry allowed measuring proper motions of dwarfs with the Hubble Space Telescope \citep{sohn2020hst} and Gaia \citep{gaia_collab,fritz2018gaia,pace2019proper} and have expanded the possibilities of analyses greatly in the past few years and promise to continue doing so in the future \citep{kallivayalil2015hubble}. However, reconstructed satellite orbits are necessarily predicated on some model of the MW, and the time range over which the orbit integration remains close to the true orbit is determined by the ability of the model to resemble the true potential \citep{gomez2015and}. The positions of the centers of the clumps of stream stars in such a model, defined by the mean orbit of the group of stars forming each stream, should also be stable with time. Studying this action-space invariance can therefore be used to quantify how close a potential model is to the true potential in terms of how well it preserves two important  properties of the system: the property that stars in the same stream are on neighboring orbits, quantified by the tightness of their action-space clump, and the stability of the mean orbit about which the stars scatter, quantified by the stability of the location of the clump in action space with time.

In this paper, we analyze streams in their action-space properties of streams in the MW-like cosmological simulations with different accretion histories from the Feedback In Realistic Environments (FIRE-2) suite \citep{hopkins2018} described in (Sec.~\ref{subsec:sims}). In Sec.~\ref{sec:potential_model} we list different potential models utilized and how well they preserve action-space information (Sec.~\ref{sec:action}). We specifically examine the ability of the multipole expansion (Sec.~\ref{subsec:Mult_pow_spec}) to model galaxies with different merger histories, in order to understand to what extent these models can successfully produce a coherent, and consistent, action-space distribution for stellar streams. We use statistical measures of information detailed in Sec.~\ref{Sec:KLD} on our entire sample to quantify the degree of clustering and stability of action space for different potential models in Sec.~\ref{subsec:uniformKLD} and \ref{subsec:neighborKLD} respectively. We present our results in Sec.~\ref{Sec:Conc}.

\section{Action space of Tidal Streams in Cosmological simulations}\label{method}

In this section, we introduce the framework used to compare potential models in terms of the degree of action space clustering and coherence for stars in stellar streams. We start with a brief description of the simulations utilized for the analysis (\S\ref{subsec:sims}--\ref{sec:centering}) and introduce the basis expansion used for the potential models (\S\ref{sec:potential_model}). Finally, we review how actions are calculated (\S\ref{sec:action}) and describe the statistics used to quantify the clustering and stability of the action space when assuming various potential models (\S\ref{Sec:KLD}). 

\subsection{Simulations}
\label{subsec:sims}
We study cosmological zoom-in baryonic simulations of MW-mass galaxies from the Latte suite \citep[introduced in][]{wetzel2016reconciling}, which is part of the FIRE project.\footnote{\url{http://fire.northwestern.edu/latte}} Simulations were run using the Gizmo code \citep{hopkins2015new}, which utilizes a TREE+PM gravity solver, a Lagrangian meshless finite mass (MFM) solver for hydrodynamics that provides adaptive spatial resolution. The simulations in this paper were run with the FIRE-2 physics model \citep{hopkins2018}, which implements star formation and stellar feedback parameters from realistic stellar evolution models (STARBURST99; \citealt{leitherer1999starburst99}), consistent with the Lambda cold dark matter cosmology from the Planck \citet{collaboration2015planck}. \cite{wetzel2016reconciling} and \cite{hopkins2018} give an in-depth description of the simulations from the FIRE-2 suite. We select three galaxies from the Latte suite, referred to in this paper as \mi{}, \mf{}, and \mw{}, for our analysis that are similar to MW in their stellar and gas mass. \mi{} and \mf{} are similar to MW in their stellar morphology while \mw{} is less disky \citep{hopkins2018, sanderson2020synthetic}. In the Latte suite, we chose halos agnostic to their formation history, so the sample is representative of varying merger histories, until the present day (redshift \textit{z} = 0) \citep{samuel2021planes}. These halos have a total mass of $1--2 \times 10^{12}$ M\textsubscript{\(\odot\)} with average mass of baryons $m_\mathrm{b} = 7100$ M\textsubscript{\(\odot\)} and DM particle $m_\mathrm{DM} = 35,000$ M\textsubscript{\(\odot\)}. Table~\ref{tab:merger} lists radial and tangential velocities of the most massive merging satellite at the first pericenter time $t_\mathrm{peri}$ where the satellite is $d_\mathrm{peri}$ from the center of the main halo, along with the mass ratios of the mergers observed in the simulations measured at $t_\mathrm{peri}$, computed as,

\begin{itemize}
\item {Infall mass ratio (IMR): The mass of the main halo divided by mass of the satellite just before infall.}
\item Total mass ratio (TMR): The mass of the main halo divided by the mass of the merging satellite.
\item Pericenter mass ratio (PMR): The mass of the main halo enclosed within the $d_\mathrm{peri}$ divided by the mass {brought in by} the merging satellite {enclosed within $d_\mathrm{peri}$ from the center of the satellite}. 
\end{itemize}

% ##################################
% Merger info table
% ##################################
\begin{table*}
\centering
\caption{Characteristics of potent mergers in Simulations}
\resizebox{\textwidth}{!}{
\begin{tabular}{c|ccccccc}
\multicolumn{1}{l}{\textbf{Simulation}} & \multicolumn{1}{l}{\textbf{\begin{tabular}[c]{@{}l@{}} $\mathbf{t_{peri}}$ \\ {(}Gyr{)}\end{tabular}}} & \multicolumn{1}{l}{\textbf{\begin{tabular}[c]{@{}l@{}} $\mathbf{d_{\mathrm{peri}}}$ \\ {(}kpc{)}\end{tabular}}} & \multicolumn{1}{l}{\textbf{IMR}} & \multicolumn{1}{l}{\textbf{TMR}} & \multicolumn{1}{l}{\textbf{PMR}} & \multicolumn{1}{l}{\textbf{\begin{tabular}[c]{@{}l@{}}$\mathbf{v_{\mathrm{rad}}}$ \\ {(}km s$^{-1}${)}\end{tabular}}} & \multicolumn{1}{l}{\textbf{\begin{tabular}[c]{@{}l@{}}$\mathbf{v_{\mathrm{tan}}}$ \\ {(}km s$^{-1}${)}\end{tabular}}} \\ \hline \hline
\mi{} & 7.95 & 29.53 & {14.6} & 45.5 & {17.4}  & -13.36 & 290.24 \\ \hline
\textsf{m12f -1} & 10.8 & 35.74 & {7.1} & 15.9 & {7.9}  & 6.22  & 362.52 \\
\textsf{m12f - 2} & 7.2 & 5 & {8.9} & 10.25 & {4.6} & -43.98  & 263.76 \\ \hline
\mw{} & 8 & 7.67 & {7.1} & 8.3 & {2.6}  & 106.82 & 242.11 \\ \hline
\end{tabular}
}\\

\textbf{Note.} $t_{\mathrm{peri}}$: Time of closest approach between the main galaxy and the satellite. $d_{\mathrm{peri}}$: Distance between satellite and main galaxy at closest approach. {IMR: Mass Ratio at infall, $M_{\mathrm{main}}/M{\mathrm{sat}}$ just before infall,} TMR: Total Mass Ratio, $M_{\mathrm{main}}/M{\mathrm{sat}}$ at $t_\mathrm{peri}$, and PMR: Pericenter Mass Ratio, $M_{\mathrm{main}}(<d_{\mathrm{peri}})/M_{\mathrm{sat}}(<d_{\mathrm{peri} \mathrm{\, from \, sat}})$ evaluated at $t_{\mathrm{peri}}$. $v_{\mathrm{rad}}$, $v_{\mathrm{tan}}$: radial and tangential velocities of the satellite with respect to the main galaxy. The non-zero $v_{\mathrm{rad}}$ is a limitation of finite time resolution but regardless comments on how radial the merger is.
\label{tab:merger}
\end{table*}

\mi{} lacks any major infalling satellites after 7 Gyr. \mi{}'s largest merger has a TMR of 45.5 and a slow radial velocity $v_\mathrm{rad}$, making it a rather weak interaction. This isolated halo acts as a baseline to study the adverse effects of mergers on potential modeling. \mf{} has two major mergers at 10.8 \textrm{Gyr} (m12f-1) and 7.2 \textrm{Gyr} (m12f-2) with TMRs 15.9 and 10.25, and PMRs 7.9 and 4.6, respectively. m12f-1 merger is comparable to the MW's interaction with the progenitor of the Sagittarius stream, Sgr dSph galaxy \citep{jiang2000orbit,niederste2010re,de2015star,gibbons2017tail} in terms of merger ratios with the distance at the first pericenter 35.7 \textrm{kpc}. In case of \mw{}, the TMR is 8 and PMR is close to {2.6} similar to MW's merger with LMC which has an estimated TMR of 7.6--10 in the broadest confidence interval \citep{penarrubia2015_LMC,shipp2021measuring,vasiliev2021tango} with significantly higher radial velocity 106.8 \textrm{km/s} and the pericenter distance 7.2 \textrm{kpc}, which is not the case for the LMC \citep{besla2007magellanic}.

\subsection{Stream identification}\label{sec:stream_ind}
Stellar streams in these halos are selected based on the number of star particles (between $120$ and $10^5$ particles), pairwise separation between them (maximum value of separation between any two star particles is greater than 120 kpc), the local velocity dispersion \citep{panithanpaisal2021galaxy}, and bound to host between 2.7-6.5 Gyr ago. Since each star particle has a mass of $\sim 5000 M\textsubscript{\(\odot\)}$, streams have a lower mass limit of $10^6 M\textsubscript{\(\odot\)}$. We know precisely which stars belong to each stream {[we have linked the stars to their DM halos in the merger tree \citep{panithanpaisal2021galaxy}]}. The number of streams in each of the simulations are different, while \mi{} and \mf{} have nine and eight streams respectively, \mw{} only has two resolved streams. We track the formation and evolution of streams in time \citep{panithanpaisal2021galaxy} using the halo catalogs \citep{Behroozi_2012a} connected in time using consistent trees \citep{Behroozi_2012b}.   

\subsection{Centering}\label{sec:centering}

\begin{figure}
    \includegraphics[width=\linewidth]{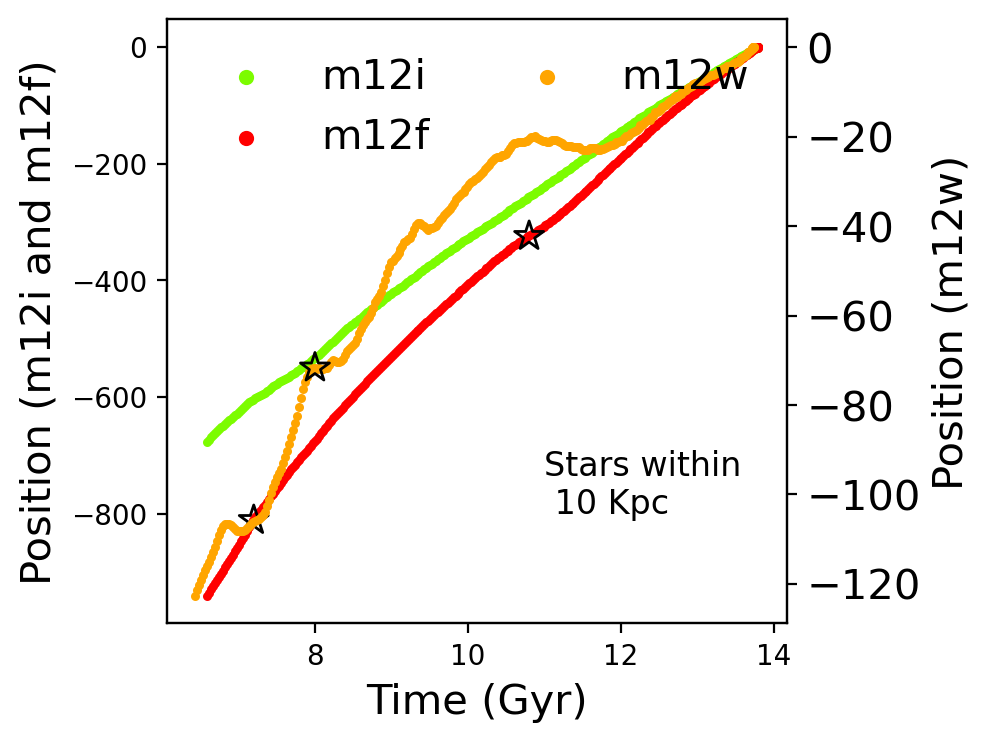}
    
    \caption{The dynamically evolving COM in simulation coordinates for \mi{} (green), \mw{} (orange), and \mf{} (red). The re-centering is done with respect to the present-day positions. While \mi{} has smooth COM positions with time, accelerations in COM are prominent in \mf{} and \mw{} at $t_\mathrm{peri}$ (black star). This is due to the infalling satellites displacing the barycenter position and velocity (reflex motion) during the merger. This effect is stronger in \mw{} with a more massive satellite and lower $d_\mathrm{peri}$.}
    \label{fig:COM_gal}
\end{figure}

The simulations are run in an arbitrary simulation box frame, where the total momentum of the system is nonzero. We first establish a galactocentric coordinate system (\textbf{x},\textbf{v}) for each snapshot aligned with stars in the galaxy. We compute the central position of each halo by calculating the center of mass (COM) of stars within a sphere and iteratively shrinking the sphere and recomputing the COM until a consistent central location is reached. Next, we estimate the central velocity {using the mean velocities of} stars within 10 kpc of the center. We pick this location as (\textbf{x},\textbf{v}) = (0,0) and rotate all the snapshots of a simulation such that the Galactic disc is planar in \textit{xy} at the present day, this is a good approximation as the angular momentum of the disk does not vary for the second half of the simulation when the star formation is steady rather than bursty (happens usually at z = 0.5) \citep{garrison2018origin,santistevan2021origin,gurvich2022rapid}. 
Fig.~\ref{fig:COM_gal} plots the total COM displacement in simulation coordinates over a period of 7 Gyr. Note the accelerating COM positions in the case of \mf{} (red) and \mw{} (orange) due to ongoing mergers at respective $t_\mathrm{peri}$ (marked with a star). The infalling satellite in \mw{} accelerates the main halo significantly during the interaction, leading to a displacement in both positions and velocities (reflex motion). As such, the COM of the outer regions are significantly displaced from the Galactic disk due to this reflex motion \citep{erkal2019total, cunningham2020quantifying,petersen2020reflex,garavito2021LMCimpact}. 

%##########################################################
%Figure : COM
%#########################################################

\begin{figure}
    \includegraphics[width=\linewidth]{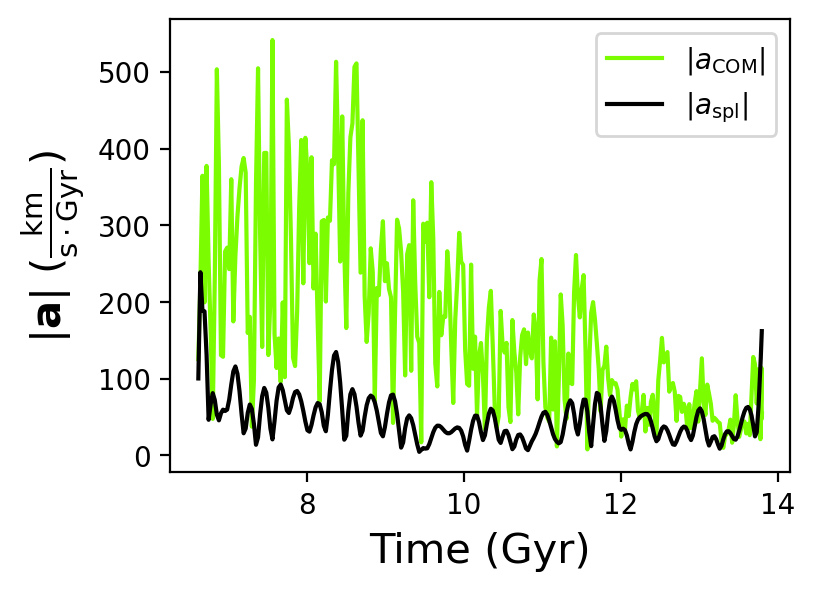}
    \caption{{The total magnitude of COM acceleration (green) along with the spline accelerations (black) calculated using the smooth cubic spline fits for each coordinate axis for \mi{}. The actual acceleration is jittery, highly oscillatory, and large while the spline values are significantly smooth and lower in comparison. The spline fit COM positions and velocities are overall better suited for potential modeling.}}
    \label{fig:acc_example}
\end{figure}

There are also high-frequency jitters in the COM position due to active star formation near the center \citep{orr2021fiery}, This jerky COM motion causes fluctuations in the central accelerations (green points in Fig.~\ref{fig:acc_example}) that lead to a non-conserved reference frame. {These jittery accelerations are insensitive to the centering method and the aperture used to find the central positions and velocities.} We limit this non-conserved nature of the axis frame caused by jittery COM by fitting three cubic approximations (one for each axis) on the COM positions over time ensuring that the new functions are doubly differentiable. These expansions act as smooth proxies for the original center positions and velocities and minimize the effect of shaky accelerations. Since we are using cubic spline expansions, we track the same COM position in time and the maximum difference in central velocities is 17 km/s. Fig.~\ref{fig:acc_example} plots the total magnitude of spline accelerations (black) for \mi{}, obtained via the second derivative of the spline functions, along with the magnitude of actual acceleration (green), found using numerical differentiation of real central positions for \mi{}. Note the highly erratic and larger actual accelerations in contrast with smooth and smaller spline accelerations. This methodology does not affect the potential modeling fits for the stellar streams as their orbits are far from the Galactic Center. The Cartesian accelerations along with their spline counterparts are posted in Appendix \ref{app:acc_m12i}.   

%##########################################################
%Figure : Acc_example m12i.
%#########################################################

\begin{figure}
    \includegraphics[width=\linewidth]{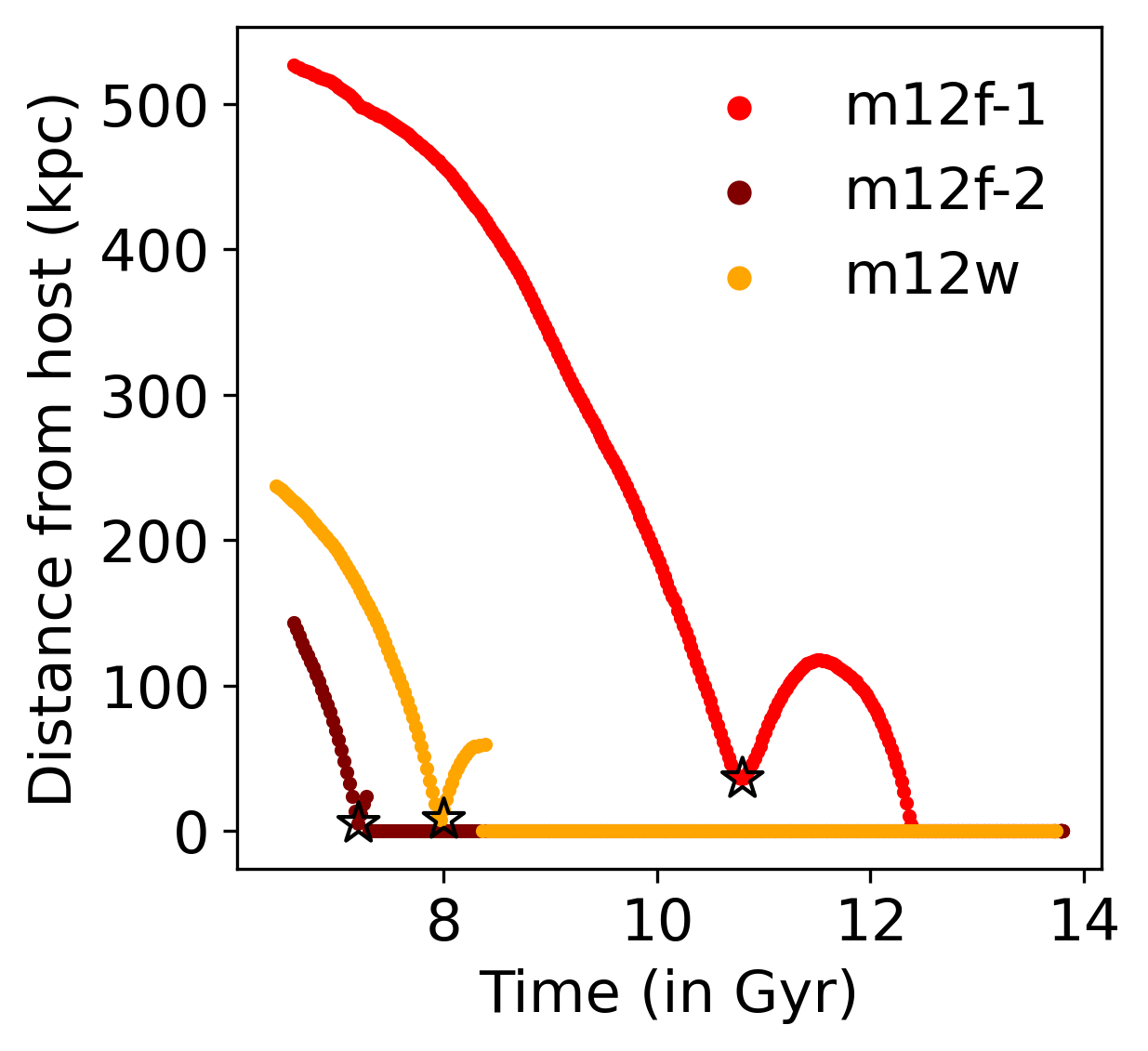}
    
    \caption{{Pericenter distance of the merging satellite with respect to the host galaxy for \mf{} and \mw{}. The mergers happen at 7.2 and 10.8 Gyr for \mf{} (two different satellites) and at 8 Gyr for \mw{}. The black stars illustrate the time at the first pericenter $t_{\textrm{peri}}$.}}
    \label{fig:LMC_m12f}
\end{figure}

We use the spline approximations to define central positions, velocities, and fix the orientation of the disk in each simulation. Fig.~\ref{fig:LMC_m12f} plots the pericenter position of the merging galaxy from the central halo for about 7 Gyr in the past. The m12f-1 merger has a prolonged infall time, while other mergers are relatively short. The shorter time scale mergers are closer to the main halo at $t_\mathrm{peri}$ (black star) and difficult to incorporate into a time-evolving potential model.    
%##########################################################
%Figure : LMC_pericenter
%#########################################################

%###############POTENTIAL MODEL###########################
%#########################################################
%#########################################################

\subsection{Potential models}\label{sec:potential_model}

We use four types of potential models: \textbf{(i). a time-independent analytic potential (TIAP)} model described by a combination of a Navarro-Frenk-White (NFW) profile for the DM halo \citep{NFW1996}, a spherical power law for the central bulge \citep{binney2011galactic} (assuming spherical symmetry) and a Miyamoto-Nagai profile to model the stellar disk \citep{MN1975} (assuming axisymmetry). These parametric potential profiles are traditionally utilized at \textit{z} = 0 fitted to observational data \citep[e.g][]{McMillan_anly,Bovy_2014,mcmillan2016mass} or a simulation snapshot \citep{Buist_2014,sanderson2017modeling}. \textbf{ (ii). a semi-time independent analytic potential (STIAP)} model that fixes the shape of the potential using TIAP but accounts for the changing mass with a linear mass ratio scaling at each time. \textbf{(iii). a time-evolving multipole potential (TEMP)} using BFEs, a low-order spherical harmonic and azimuthal harmonic basis expansion fit under axisymmetry conditions for DM and baryons, respectively, evaluated at each time snapshot independently, and \textbf{(iv). a semi-time independent multipole potential (STIMP)} model calculated at \textit{z} = 0 using the same expansions and symmetrical structure as TEMP with a linear mass ratio scaling to account for the changing mass of the halo in time. We fit these different potential models using the galaxy dynamics package Agama \citep{agama2018}.

\subsubsection{Analytic models}\label{Sec:anly_model}
We fit a parametric potential model for the DM halo, Galactic disk, and the central bulge of the galaxy independently at \textit{z} = 0. We use the NFW profile to model the DM halo and hot gas ($T_\mathrm{gas} \geq 10^{4.5} \mathrm{K}$) described by
\begin{equation}
    \Phi_\mathrm{NFW}(r)=-\frac{M_\mathrm{NFW}}{r} \ln \left(1+\frac{r}{r_s}\right)
\end{equation} \label{NFW}

where $\Phi_\mathrm{NFW}(r)$ is the potential at spherical radius $r$, $M_\mathrm{NFW}$ is the mass enclosed {in a radius of $\sim 5.3 r_s$} and $r_s$ is the scale radius of the main halo. The NFW profile is parameterized using $M_\mathrm{NFW}$ and $r_s$ because of the sensitivity of the stellar orbits to the locally enclosed mass. We fit the total density of each DM halo using the density form of the NFW profile for DM particles within 300 kpc.  

We model the Galactic Disk defined by the star particles and cold gas ($T_\mathrm{gas} < 10^{4.5} \mathrm{K}$) within 5 kpc $\leq$ cylindrical radius (\textit{R}) $\leq$ 30 kpc and height $|Z| \geq 3$ kpc from the COM using the Miyamoto-Nagai (MN) profile in azimuthal coordinates, given by 
\begin{equation}
    \Phi_\mathrm{MN}(R,Z)=-\frac{M_\mathrm{MN}}{\sqrt{R^{2}+(a+\sqrt{Z^{2}+b^{2}})^{2}}} .   
\end{equation}\label{MyNa}

We fit the 2D density profile of the simulation with the density functional form of MN profile using the scale mass ($M_\mathrm{MN}$), scale radius ($a$) and scale height ($b$) for this Galactic disk.  

The central bulge of stars and cold gas within $r\leq 3$ kpc of the COM is modeled using a spherical power-law-based density profile: 
\begin{equation}
\rho(r)=\rho_{0}\left(\frac{{r}}{r_o}\right)^{-\gamma}\left[1+\left(\frac{{r}}{r_o}\right)^{\alpha}\right]^{\frac{\gamma-\beta}{\alpha}}
\times \exp \left[-\left(\frac{{r}}{r_{\mathrm{cut}}}\right)^{\xi}\right]
\end{equation}

where $r$ is the spherical radius, $\rho_0$ is the normalized density, $r_\mathrm{cut}$ is the outer cutoff radius, and $\alpha, \gamma, \beta$, and $\xi$ are the power parameters. We again fit the density of the spherical bulge. For all our fits, we utilize a Nelder-Mead optimizer \citep{gao2012} {to minimize differences in log-density (i.e., relative error)}. Table ~\ref{tab:ana} lists the TIAP fits for the halos. 

\begin{table*}
\caption{Best-fit parameters for the Analytic Potential Model computed at \textit{z} = 0 for the Three Simulations}
\resizebox{\textwidth}{!}{
\begin{tabular}{c|cccccccccccc}
\hline
\textbf{Simulation} & \textbf{\begin{tabular}[c]{@{}c@{}}$\textrm{M}_\textrm{NFW}$\\ {[}$10^{10} \textrm{M}_{\odot}${]}\end{tabular}} & \textbf{\begin{tabular}[c]{@{}c@{}}$r_s$\\ {[}kpc{]}\end{tabular}} & \textbf{\begin{tabular}[c]{@{}c@{}}$\textrm{M}_\textrm{MN}$\\ {[}$10^{10} \textrm{M}_{\odot}${]}\end{tabular}} & \textbf{\begin{tabular}[c]{@{}c@{}}a\\ {[}kpc{]}\end{tabular}} & \textbf{\begin{tabular}[c]{@{}c@{}}b\\ {[}kpc{]}\end{tabular}} & \textbf{\begin{tabular}[c]{@{}c@{}}$\rho_o$\\ {[}$\textrm{M}_{\odot}/ \textrm{kpc}^3${]}\end{tabular}} & \textbf{\begin{tabular}[c]{@{}c@{}}$r_o$\\ {[}kpc{]}\end{tabular}} & \textbf{$\alpha$} & \textbf{$\beta$} & \textbf{$\gamma$} & \textbf{$\xi$} & \textbf{\begin{tabular}[c]{@{}c@{}}$r_\textrm{cut}$\\ {[}kpc{]}\end{tabular}} \\ \hline \hline
\mi{} & 45 & 12.3 & 4.49 & 2.47 & 0.85 & $5.9 \times 10^{9}$ & 0.28 & 1.81 & 1.61 & 0 & 2 & 4 \\ \hline
\mf{} & 65 & 14.1 & 5.14 & 3.11 & 0.81 & $4.8 \times 10^{9}$ & 1.93 & 1.06 & 6.70 & 0.5 & 2 & 4 \\ \hline
\mw{} & 39 & 8.91 & 4.01 & 2.46 & 0.84 & $8.5 \times 10^{9}$ & 2.54 & 2.30 & 1.43 & 0.5 & 2 & 4 \\ \hline
\end{tabular}}\\

\textbf{Note}. Note we perform least square optimization to find the best fit using the Nelder-Mead method described by \citep{gao2012}.

\label{tab:ana}
\end{table*}

Given the TIAP model, we from the STIAP by extrapolating the TIAP at other times with a mass ratio scaling, between the mass of the halo at each time step with respect to the total mass of the halo at the present day.

\subsubsection{Low-order Multipole model}\label{Sec:mult_model}

The multipole model uses a combination of BFEs to accurately describe the density and potential. These potential expansions are found by replacing two of the three integrals from the Green's function solution to the Poisson equation, with a sum in some BFEs \citep[e.g][]{lowing_BFE, garavito2021LMCimpact}. We use a spherical harmonic basis to describe the DM halo and hot gas ($T_\mathrm{gas} \geq 10^{4.5} \mathrm{K}$) components of the potential and an azimuthal harmonic expansion for the stellar and cold gas ($T_\mathrm{gas} < 10^{4.5} \mathrm{K}$) components.  

To model the DM and hot gas, we use a spherical-harmonic expansion:
\begin{equation}
    \Phi(r, \theta, \phi)=\sum_{\ell, m} \Phi_{\ell, m}(r) Y_{\ell}^{m}(\theta, \phi)
\label{eq:mult_pot}
\end{equation}

where ($r, \phi, \theta$) are spherical coordinates in the centered frame, $Y_\ell^m$ are the spherical harmonics for the multipole term $\ell$, and $-\ell\leq m \leq \ell$. $\Phi_{\ell,m}(r)$ is the amplitude:

\begin{equation}
    \begin{array}{r}
\Phi_{\ell,m}(r)=-\frac{4 \pi G}{2 \ell+1}\left[r^{-1-\ell} \int_{0}^{r} d r^{\prime} \rho_{\ell m}\left(r^{\prime}\right) r^{\prime \ell+2}+\right. \\
\left.r^{\ell} \int_{r}^{\infty} d r^{\prime} \rho_{\ell m}\left(r^{\prime}\right) r^{\prime 1-l}\right] 
\end{array}
\label{eq:mult_exp}
\end{equation}

where $\rho_{l m}(r)$ can be expressed as two integrals of density at each radius, with normalization constants $A_{l m}$: 
\begin{equation}
    \rho_{l m}(r)=\int_{0}^{\pi} d \theta \int_{0}^{2 \pi} d \phi \rho(r, \theta, \phi) Y_{l}^{m}(\theta, \phi) A_{l m} .
\end{equation}

This quantity is obtained from the particle distribution using a penalized spline estimate in logarithmically spaced radii fit to spherical harmonics coefficient calculated for each particle $i$ \citep[][Appendix A.2.5]{vasiliev2018agama}:

\begin{equation}
    \rho_{l, m ; i} \equiv m_{i} \sqrt{4 \pi} \tilde{P}_{l}^{m}\left(\cos \theta_{i}\right) \operatorname{trig} m \phi_{i}, 
\end{equation}

where $\tilde{P}_{l}^{m}$ are normalized associated Legendre polynomials and 
\begin{equation*}
\operatorname{trig} m \phi \equiv\left\{\begin{aligned}
1 &, \quad m=0 \\
\sqrt{2} \cos m \phi &, \quad m>0 \\
\sqrt{2} \sin |m| \phi &, \quad m<0 .
\end{aligned}\right.
\end{equation*}

To model the stars and cold gas, we use a BFE based on sets of Fourier harmonics in azimuthal coordinates, where potential and density are

\begin{equation}
\{\rho, \Phi\}(R, Z, \phi)=\sum_{m=0}^{\infty}\{\rho, \Phi\}_{m}(R, Z) \exp (i m \phi)    
\end{equation}

where $\rho_m$ and $\phi_m$ are given by:

\begin{equation}
\Phi_{m}(R, Z)=-G \int_{-\infty}^{\infty} d Z^{\prime} \int_{0}^{\infty} d R^{\prime} \rho_{m}\left(R^{\prime}, Z^{\prime}\right) \Xi_{m}\left(R, Z, R^{\prime}, Z^{\prime}\right)
\end{equation}

and $\Xi_m$ is the Green's function in cylindrical coordinates. For the axisymmetric model utilized in this paper, only the $m = 0$ {harmonic} exists for the azimuthal basis and only $\ell$ even terms are nonzero for the spherical harmonics expansion.\footnote{We provide the low-order ($\ell_\textrm{max} = 4$ and $m_\textrm{max} = 4$) multipole potential models for the FIRE-2 suite of cosmological simulations computed at the present day under two different symmetry assumptions: axisymmetry, which can be used to compute actions in cylindrical coordinates, and with no symmetry assumptions, which is better at modeling mergers with high fidelity. These expansions and a demonstration Jupyter notebook are available at \url{https://web.sas.upenn.edu/dynamics/data/pot_models/}.} We compute our expansion coefficients  $\Phi_{l, m}(r)$ on a 1D grid of size of 25 kpc and $\Phi_{m}(R,Z)$ on a 2D meridional plane of grid size 25 kpc in $R$ and $Z$ \citep{agama2018}. 

These two expansions together account for the total mass distribution of matter and form the Galactic potential profile. For TEMP, we find these expansion coefficients at each time step in the specified coordinates form a time-evolving model. For STIMP, we use the expansion calculated at the present day and rescale the potential to the total halo mass (total mass in all species within $R_\textrm{vir}$) at each time step similar to the STIAP.    

\begin{figure*}{
    \includegraphics[width=\textwidth,height=\textheight,keepaspectratio]{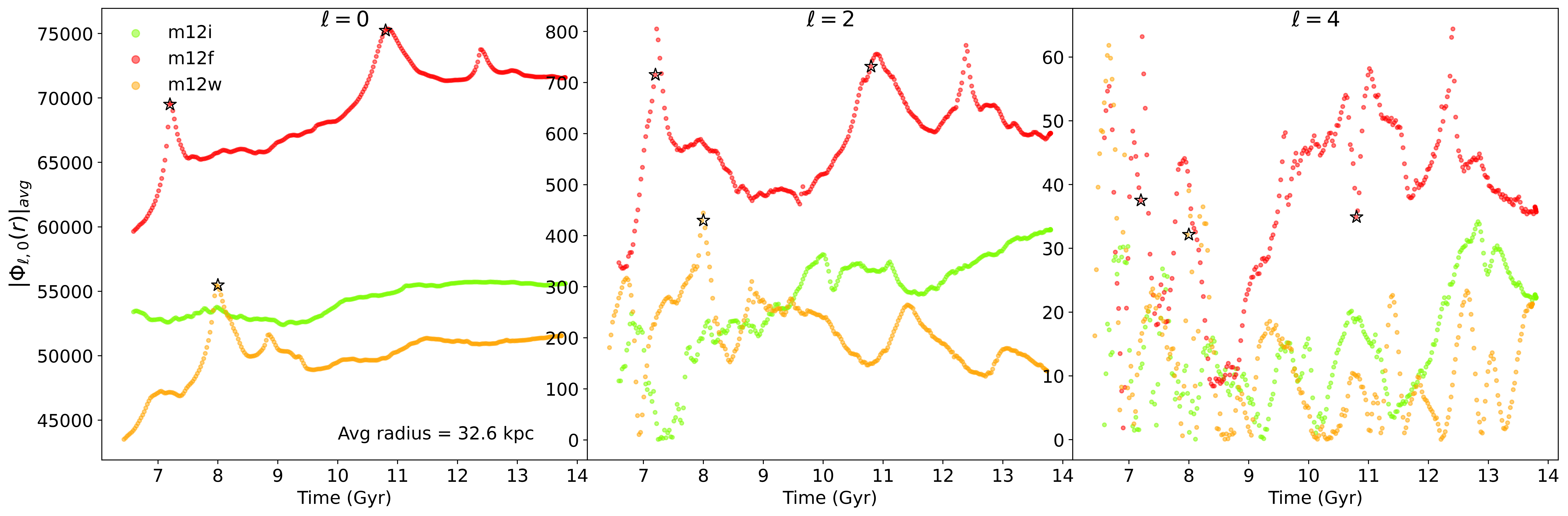}
    }
    \caption{{Amplitude in $|\Phi_{\ell,0}|_r$  averaged over radii for even poles of the multipole model with $\ell \leq 4$ described in Sec.~\ref{Sec:mult_model} for all the halos computed independently at each time step. Due to the axisymmetry assumption, the odd poles vanish and the {amplitude} is stored only in $m = 0$ for even poles. The black stars on the plot mark the $t_\mathrm{peri}$ for \mf{} and \mw{}. The increased {amplitude} near the merger demonstrate the ability of the higher poles in spherical harmonics in modeling a satellite infall.}}
    \label{fig:power_spec}
\end{figure*}

Fig.~\ref{fig:power_spec} plots the $|\Phi_{\ell,0}|$ (amplitude) averaged over the radii, computed at even poles up to $\ell \leq 4$ of TEMP (the odd poles are zero due to axisymmetry). The increased {amplitude} in higher poles around the merger time (marked by black stars for \mf{} and \mw{}) implies an increase in total mass at the pericenter passage of the satellite as the merger causes the potential to deepen. The higher poles have limited ability to capture potential perturbations caused by the merging satellite even under axisymmetry assumptions. 

%##########################################################
%Figure : Power spectrum_ avg rad.
%#########################################################

\begin{figure}{
    \includegraphics[width=\linewidth]{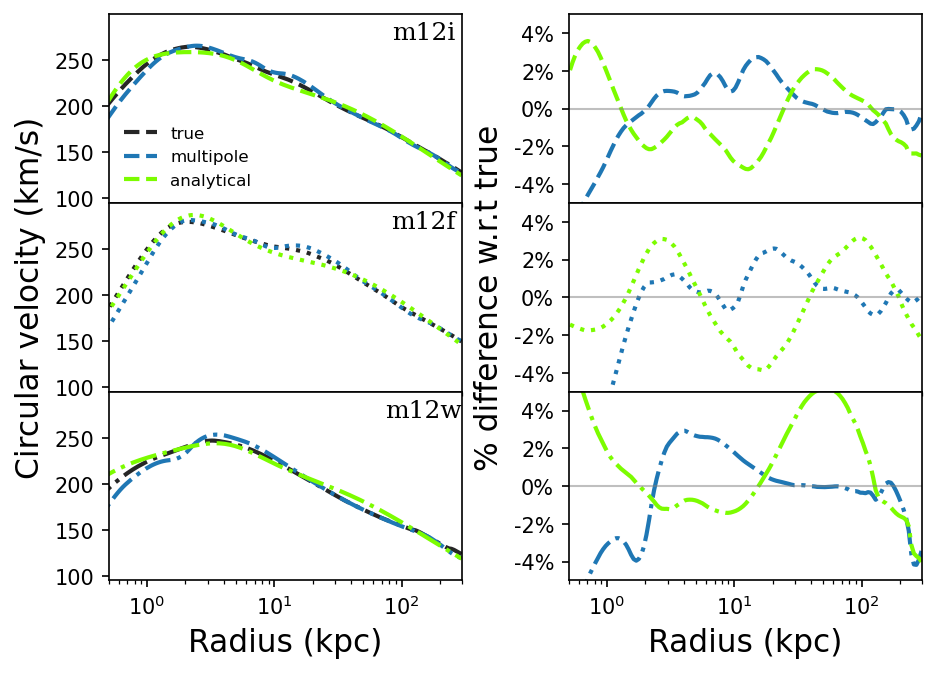}
    }
    \caption{{Rotation velocity curve (left) computed using different potential models at present day along with the true rotation curve (black). Note that both multipole (blue) and analytic (green) models produce rotation curves consistent with the true curves (accuracy of $\pm5\%$ at all radii (right)) of the true rotation for \mi{} (top), \mf{} (middle), and \mw{} (bottom). This serves as a preliminary check for our models.}}
    \label{fig:rot_curve}
\end{figure}

We plot the rotation curves for the potential models along with the true rotational velocities (left) for our simulations in Fig.~~\ref{fig:rot_curve} at \textit{z} = 0 for \mi{} (top), \mf{} (middle), and \mw{} (bottom). All the potential models at the present day fit the actual rotation curve well within $\pm5\%$ (percentage difference plots on the right). We fine-tune the TIAP parameters to improve the velocity fits under the values given in Table~\ref{tab:ana}, but fine-tuning the rotation curves does not significantly change the parameters presented. 
%##########################################################
%Figure : Rotation curve.
%#########################################################

To summarize, we consider four potential models. 
\begin{itemize}
    \item \textbf{TEMP}: in which we compute the multipole expansion at each snapshot independent of the others
    
    \item \textbf{STIMP}: where the multipole expansion is computed at present day and used for all time steps with a mass scaling hence the shape of the potential remains the same, while the mass evolves with time 
    
    \item \textbf{TIAP}: where the potentials are fit at present day and utilized at each time step
    
    \item \textbf{STIAP}: that is TIAP used with a mass ratio scaling for at different time steps.
\end{itemize}

We list the potential models along with their shorthand and labels used throughout this paper in Table~\ref{tab:pot_models}. Both the semi-time-independent and time-independent models are henceforth labeled time-independent models (TI) due to fixed potential shape.     

\begin{table}
\caption{Potential Models used in this paper and their time dependence} 
\footnotesize
% \resizebox{\textwidth}{!}
{\begin{tabular}{lll}
\textbf{Model}      & \textbf{Time} & \textbf{Name} \\ \hline
Multipole potential & Evolving      & TEMP          \\
Multipole potential & Semi-independent (\textit{z} = 0) & STIMP    \\
Analytic potential  & Semi-independent (\textit{z} = 0) & STIAP    \\
Analytic potential  & Independent (\textit{z} = 0) & TIAP \\ \hline 
\end{tabular}}\\

\textbf{Note.} Note that Name is the abbreviation for the corresponding model used throughout the paper.The time independent potential has a fixed shape and mass at all time steps, while the semi-independent potentials only have a constraint on the shape with evolving mass scaling.
\label{tab:pot_models}
\end{table}

\subsection{Computing action space distributions}\label{sec:action}

In order to analyze the stellar streams accreted in the main halo, we work in their three-dimensional action space \textbf{\textit{J}}. Actions can be calculated for independent stars as long as they are in a bound orbit in an integrable potential. Then the actions are related by a canonical transformation from the six-dimensional phase space consisting of position (\textbf{\textit{x}}) and momentum (\textbf{\textit{p}}) given by :

\begin{equation}
    J_{i}=\oint p_{i} d x_{i}
\end{equation}

for the $i^\textrm{th}$ component of \textbf{\textit{J}}, where $x_i$ are the separable coordinates and $p_i$ their conjugate momenta. Analytic expressions for actions are found for some potentials by solving Hamilton-Jacobi equations with the separation of variables \citep{binney2011galactic}. For our axisymmetric conditions, we compute actions (\textbf{\textit{$J_r$,$J_z$,$J_\phi$}}) in cylindrical coordinates computed using St\"{a}ckel fudge approximation \citep{binney2012actions}. $J_\phi$ is the total angular momentum $L_Z$ given by $(\textbf{\textit{x}} \times \textbf{\textit{p}})_Z$, and hence, independent of the potential model utilized while the other two actions are potential dependent. These actions are adiabatically invariant {in a slowly evolving potential} and should therefore show high clustering and stability at all times for independent streams in simulations where the potential is adiabatic. Clumping occurs because these stars are formed in a small phase space of localized dwarf galaxies or globular clusters before tidally stripping and forming streams \citep{helmi1999building,sanderson2015action} and the action space retains phase-space information of a stream progenitor under the absence of a potential varying merger. As these actions strongly rely on the Galactic potential, action space coherence of stellar streams is an excellent probe to identify the true potential of a galaxy \citep{sanderson2015action}. In other words, the model that produces the highest clustering and stable action space throughout the history of a stellar streams is closest to the true potential of a galaxy and hence the best model. Ideally, a good model should also be able to accurately model the actions during mergers even though satellite interactions affect the DM distribution and perturb the potential \citep{garavito2019hunting, garavito2021LMCimpact}. \cite{lilleengen2019accreted} tested whether they could constrain the gravitational potential of an Auriga halo with a similar method. They fit a potential to the simulated galaxy, varied one potential parameter, and calculated the spread in actions of an accreted group of star particles, expecting the minima of the standard deviation to be at the true parameter. This method did not reproduce the fitted potential and the actions of the star particles were not constant  

\begin{figure*}
\includegraphics[width=\textwidth,]{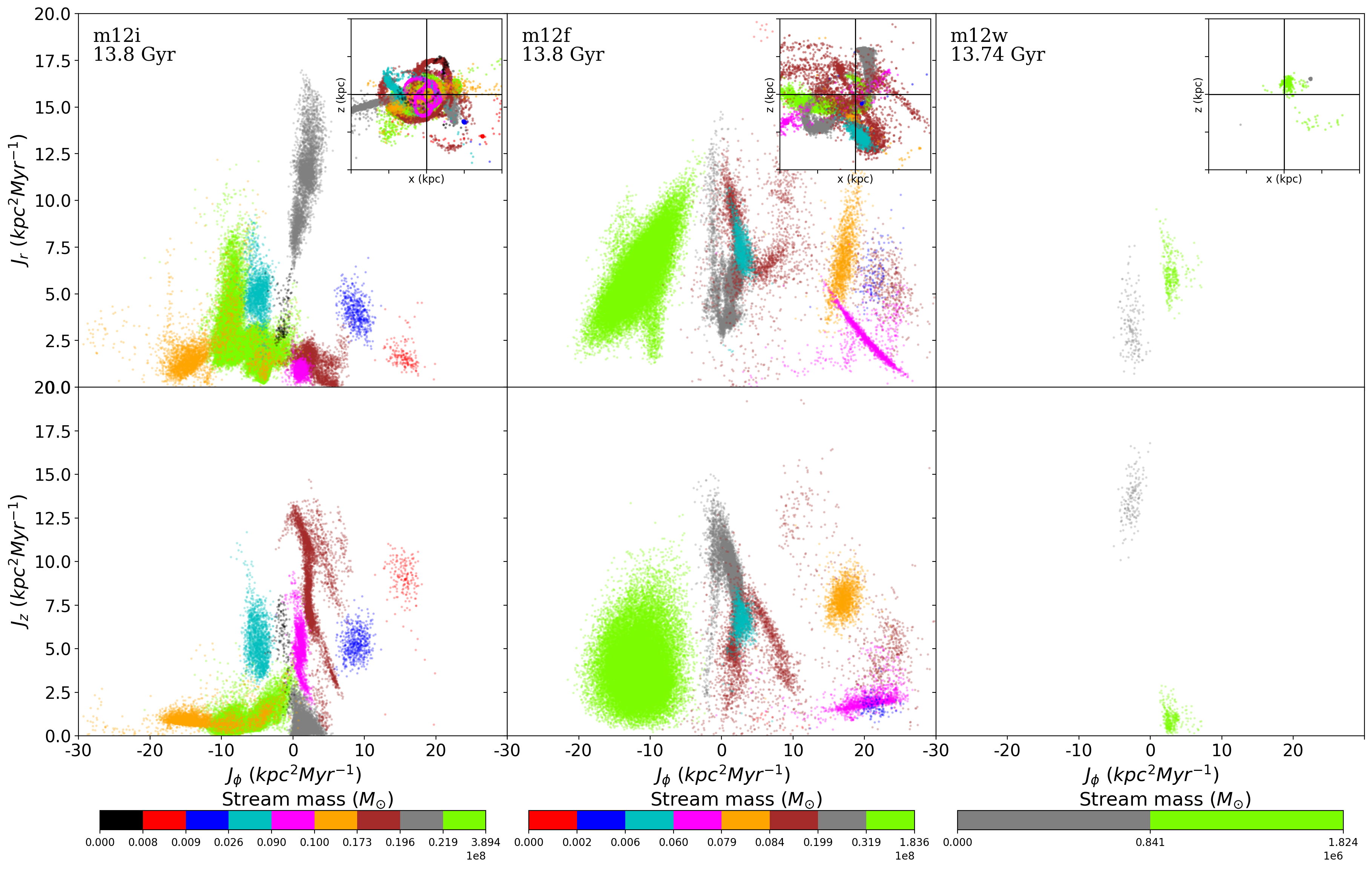}%
\caption{{Known stellar streams in action space ($J_r$-$J_\phi$ and $J_z$-$J_\phi$) for \mi{} (left), \mf{} (center), and \mw{} (right) at the present day computed using an order 4 multipole potential expansion assuming axisymmetry (TEMP). The top right corner in the first row plots the positions (\textit{Z}--\textit{X}) for these streams. Note the clustering of streams in the action space as compared to their position space. The color bar at the bottom lists the mass of each stream in solar masses starting from lowest to highest. A higher mass order in the streams leads to less clustering in the action space. Also, the different number of streams in each simulation, \mi{} and \mf{} have nine and eight streams respectively, \mw{} only has two.}}\label{fig:samp_spacel}
\end{figure*}

 In this paper, We compute the actions for all star particles in each stream for a time period of 7 -- 13.8 Gyr (301 time snapshots) for the four models using particles from the actual \textit{N}-body simulation and their true phase-space information at each time step. The alternative approach would be to take the initial conditions of particles from a stellar stream and integrate their orbit in time. Although this approach is far more complex, as reproducing trajectories in a smooth potential model ignores smaller scale perturbations on the orbits. Fig.~\ref{fig:samp_spacel} plots the stellar streams in action space (\textbf{\textit{$J_r$}}-\textbf{\textit{$J_\phi$}} and \textbf{\textit{$J_z$}}-\textbf{\textit{$J_\phi$}}) for \mi{} (left), \mf{} (center), and \mw{} (right) along with the position space ($Z$-$X$) in insets (top right corner in first row) at present day for the TEMP model discussed in Sec.~\ref{sec:potential_model}. Actions are highly clustered for each stream individually for both \mi{} and \mf{}. The stream colors correspond to the total stellar mass as shown in the color bar in Fig.~~\ref{fig:samp_spacel}. We note that more massive streams (green in Fig.~~\ref{fig:samp_spacel}) have a wider spread in action space. The few streams present in \mw{} are less massive compared to the other simulations and lack enough star particles to comment on the action space coherence.

%##########################################################
%Figure : Samp space (all)
%#########################################################

Some of the bound star particles considered within the potential models can have unbound orbits (total energy greater than zero), which would not have defined actions. Therefore, the fraction of stars with defined actions should also be maximized for each potential model.

\section{action space coherence: The Kullback-Leibler Divergence} \label{Sec:KLD}
We first analyze the action space of stellar streams for our simulations visually, to identify if the actions are clustered and stable for our models. To statistically quantify the degree of clustering and stability of the action space in a time series analysis we utilize the Kullback-Leibler divergence (KLD) \citep{kullback1951information, sanderson2015action}. KLD is a statistic that compares how similar two probability distributions are for continuous random variables. The KLD from a distribution $p(x)$ to $q(x)$ is given by

\begin{equation}
    {\mathrm{KLD}} (p: q) \equiv \int p(\mathbf{x}) \log \frac{p(\mathbf{x})}{q(\mathbf{x})} d \mathbf{x}
\end{equation}

where $p(x)$ and $q(x)$ are two independent probability distributions. Note that this statistic violates commutativity, i.e., $\mathrm{KLD} (p: q) \neq \mathrm{KLD} (q: p)$, which should be taken into account when comparing distributions. In our case, the distribution is the action space and since it is discrete and dependent on the number of stars, we estimate KLD via Monte Carlo integration to convert the above integral to a sum. Since  we know both the membership information of stars and their actions \textbf{\textit{J}} calculated using a potential $\Phi$, we compute the KLDs individually for each stream for the action space \textbf{\textit{J}} ($J_r$,$J_\phi$,$J_z$) distribution and add them to give 

\begin{equation}
    \mathrm{KLD}(\Phi)=\left.\sum_{j}^{N_{\mathrm{s}}} \sum_{i}^{N_{j}} \frac{1}{N_{j}} \log \frac{P(\boldsymbol{J} )}{Q(\boldsymbol{J})}\right|_{\boldsymbol{J}=\boldsymbol{J}_{\Phi}^{i j}}\label{eq:KLD}
\end{equation}

where $N_j$ is the number of stars in the $j^\textrm{th}$ stream, $N_s$ is the total number of streams at that time snapshot. $P(\textbf{J})$ is the probability distribution of the 3D action space of the $j^\textrm{th}$ stream, while $Q(\textbf{J})$ is the background distribution used for the comparison estimated using density estimators or analytic functions. It should be noted that these quantities are dynamically evolving as the number of streams increases with time and the probability distribution changes. For two identical distributions, or two discrete samples drawn from the same distribution, the KLD is 0; the higher the value of the KLD, the greater the difference between the distributions. 

In this work, we use the KLD in two ways: (i) to compare each action space distribution to a uniform distribution, which measures the degree of clustering (referred to as the uKLD), and (ii) to compare the action space distributions of neighboring snapshots in time, which quantifies the stability of the distribution over time (referred to as the nKLD. The KLD is calculated in \textit{nats}, i.e., information per star particle on a logarithmic scale. \cite{sanderson2015action} provide a more detailed background on the KLD statistic and also describe the framework of our analysis. 

To interpret these two statistics, we compare them with the amount of information present in each stream, calculated using a third version of the KLD called the mutual information (MI), which is the KLD of a distribution with the product of its marginal distributions \citep{sanderson2015action} {(refer to \S\ref{sec:MI}}). Comparing this number to the change in the KLD with time allows us to interpret the loss of information in the distribution in terms of the orbital decoherence of stars in streams. These numbers allow us to define the \textit{best} model of the potential as the one with (i) the highest degree of clustering (the highest uKLD value) and the greatest stability in action space (the lowest nKLD value) relative to the amount of information present in the streams (MI). We track the time evolution of KLD values with lookback time (i.e., the time before the present day defined as $\textrm{T}_\textrm{LB} = \textrm{T}_\textrm{snap} - \textrm{T}_\textrm{z=0}$).

\subsection{Clustering in time-evolving Action Space}\label{Sec:UKLD}
To quantify the degree of clustering in the time-evolving action space, we pick $Q(\textbf{J})$ in Eq.~\ref{eq:KLD} to be a uniform distribution $\mathcal{U}(J)$ in each action from its maximum to its minimum value across all time steps for all the simulations \citep{Stella2020}. The distribution $P(\textbf{J})$ is a function of phase space $\varpi^{ij}(t_n) \equiv \{\textbf{x}^{ij}(t_n),\textbf{v})^{ij}(t_n)\}$ at time $t_n$ for each stream and potential model ($\Phi(t_{n})$):
 
\begin{equation}
   \textrm{uKLD}(\Phi, t_n) =  \left.\sum_{j}^{N_{\mathrm{s}}} \sum_{i}^{N_{j}} \frac{1}{N_{j}} \log \frac{P\left\{\mathbf{J}[\mathbf{\varpi}(t_{n}),\Phi(t_{n})]\right\}}{\mathcal{U}(J_r)\mathcal{U}(J_\phi)\mathcal{U}(J_z)}\right|_{\boldsymbol{J}=\boldsymbol{J}_{\Phi}^{i j}}\label{eq:uKLD},
\end{equation}

where $J_r,J_{\phi}, J_z$ are the three actions in an axisymmetric potential. We evaluate the uKLD at each snapshot from 7 Gyr ago to the present day, for each different potential model in each simulation, using the same uniform distribution. We calculate the action space probability density $P(\mathbf{J})$ of each stream with a kernel density estimator (KDE) and evaluate the background uniform distribution $\mathcal{U}(J_r)\mathcal{U}(J_\phi)\mathcal{U}(J_z)$ analytically. The potential model closest to the true potential of the simulation produces the highest clustering in action space and hence will differ most from the uniform distribution, leading to a high value of the uKLD.

\subsection{Stability of Action Space with time}\label{Sec:NKLD}
We describe stability of our action space by calculating the nKLD: the KLD between the action space distributions produced by our model potentials at different times in the same simulation. nKLD compares the action space distribution of simultaneous snapshots and to present-day snapshots to quantify whether they significantly differ in time. Ideally they should remain the same.

For the time-dependent model (TEMP) we compare the action space distributions of two neighboring snapshots at times $t_{n-1}$ and $t_n$, $P\left\{\mathbf{J}[\mathbf{\varpi}(t_{n-1}),\Phi(t_{n-1})]\right\}$ and $P\left\{\mathbf{J}[\mathbf{\varpi}(t_n), \Phi( t_n)]\right\}$, where the $\mathbf{J}$ are calculated using the phase-space positions of the stars $\mathbf{\varpi}$ and the model potential $\Phi$ at each snapshot. As for the uKLD, we calculate nKLD for each stream individually, starting from the snapshot where the stream is first formed \citep{panithanpaisal2021galaxy}. Since different streams form at different times, we sum only over streams that are present in both snapshots, $N_s(t_{n})$ and calculate a weighted sum such that each stream contributes equally to the result \citep{Stella2020}: 
\begin{equation}
    \textrm{nKLD}(\Phi,t_n) = \sum_{j}^{N_{\mathrm{s}}(t_n)} \sum_{i}^{N_{j}} \frac{1}{N_{j} N_{s}(t_n)} \log \frac{P\left\{\mathbf{J}[\mathbf{\varpi}^{ij}(t_{n}),\Phi(t_{n})]\right\}}{P\left\{\mathbf{J}[\mathbf{\varpi}^{ij}(t_{n-1}), \Phi( t_{n-1})]\right\}} .\label{eq:nKLD_TEMP}
\end{equation}

For the TI models (STIMP, STIAP, and TIAP), the potential is fixed at $z=0$, so we instead compare the present-day action space distribution, $P\left\{\textbf{J}[\mathbf{\varpi}(t_0),\Phi(t_0)]\right\}$, to the other snapshots:
\begin{equation}
   \textrm{nKLD}(\Phi,t_n) = \sum_{j}^{N_{\mathrm{s}}(t_n)} \sum_{i}^{N_{j}} \frac{1}{N_{j} N_{s}(t_n)} \log \frac{P\left\{\mathbf{J}[\mathbf{\varpi}^{ij}(t_{n}),\Phi(t_{n})]\right\}}{P\left\{\textbf{J}[\mathbf{\varpi}^{ij}(t_0),\Phi(t_0)]\right\}} .\label{eq:nKLD_TF}
\end{equation}

In both cases, we use a KDE to calculate and evaluate the action space probability densities $P$. If $\Phi$ changes adiabatically, the action space distribution should be stationary, and the nKLD should be 0 \citep{sanderson2017modeling}. Hence, for the nKLD, a low nKLD corresponds to a higher stability of action space for a given potential model, indicating a consistent orbit family for the stream stars. 

\subsection{Information content of streams} \label{sec:MI}
Mutual information (MI) is the amount of information that can be obtained from one random variable by observing the other variable. Incidentally, one can think of MI as a measure of correlations between variables that is independent of the extent of distribution. Thus MI is used to establish the amount of orbital information a stream holds in it's action space. We compute the MI for each individual stream by calculating the KLD between the action space distribution of each stellar stream at present day, $P\left\{\textbf{J}[\mathbf{\varpi}^{ij}(t_0),\Phi(t_0)]\right\}$, and the product of the marginal distributions of \emph{all} streams in the simulation ($P^\textrm{shuf}\left\{\textbf{J}[\mathbf{\varpi}^{ij}(t_0),\Phi(t_0)]\right\}$), also at present day:
\begin{equation}
    \textrm{MI} = \sum_{i}^{N_{j}} \frac{1}{N_{j}} \log \frac{P\left\{\textbf{J}[\mathbf{\varpi}^{ij}(t_0),\Phi(t_0)]\right\}}{P^\textrm{shuf}\left\{\textbf{J}[\mathbf{\varpi}^{ij}(t_0),\Phi(t_0)]\right\}} \label{eq:MI}
\end{equation}
We obtain the product of marginal distributions by shuffling the actions with respect to one another, as in \citet{sanderson2015action}, and use a KDE to determine the densities of both distributions.
The MI thus describes the average amount of information stored in the action space of each stream about the global action space of all streams in a simulation. We can then compare the value of MI to the nKLD to measure the amount of decoherence in terms of the amount present in each stream. A higher nKLD than MI at any lookback time (i.e the time before the present day defined as $\textrm{T}_\textrm{LB} = \textrm{T}_\textrm{snap} - \textrm{T}_\textrm{z=0}$) 
indicates a loss of one stream's worth of information between snapshots.

\section{Analysis}\label{anly}

In this section, we discuss how accurately the multipole framework can accommodate perturbations from merging galaxies, and at what mass ratio the axisymmetry assumptions begin failing to represent the potential adequately during the merger. We define \emph{failure} as the inability to use the model to calculate approximate actions \emph{at all} for some subset of bound stars in the simulated galaxy. This is a narrow definition of model failure, since a model can easily succeed in calculating approximate actions for stars that are not even close to adiabatically invariant, and are not closely related to the properties of their orbits. To diagnose this, we then use the KLD to measure the potential model's ability to maintain adiabatic invariance in approximate actions for stellar streams, both in terms of orbit coherence (i.e. the overall position of each stream in action space) and in terms of stream coherence (i.e. the size of the clump of stream stars in action space). Using these metrics, we compare the performance of the multipole approach, either time evolving or static, to the static, parameterized potential models currently popular for modeling the MW.  

\subsection{Multipole power spectrum}\label{subsec:Mult_pow_spec}

\begin{figure*}
    \includegraphics[width=\textwidth,]{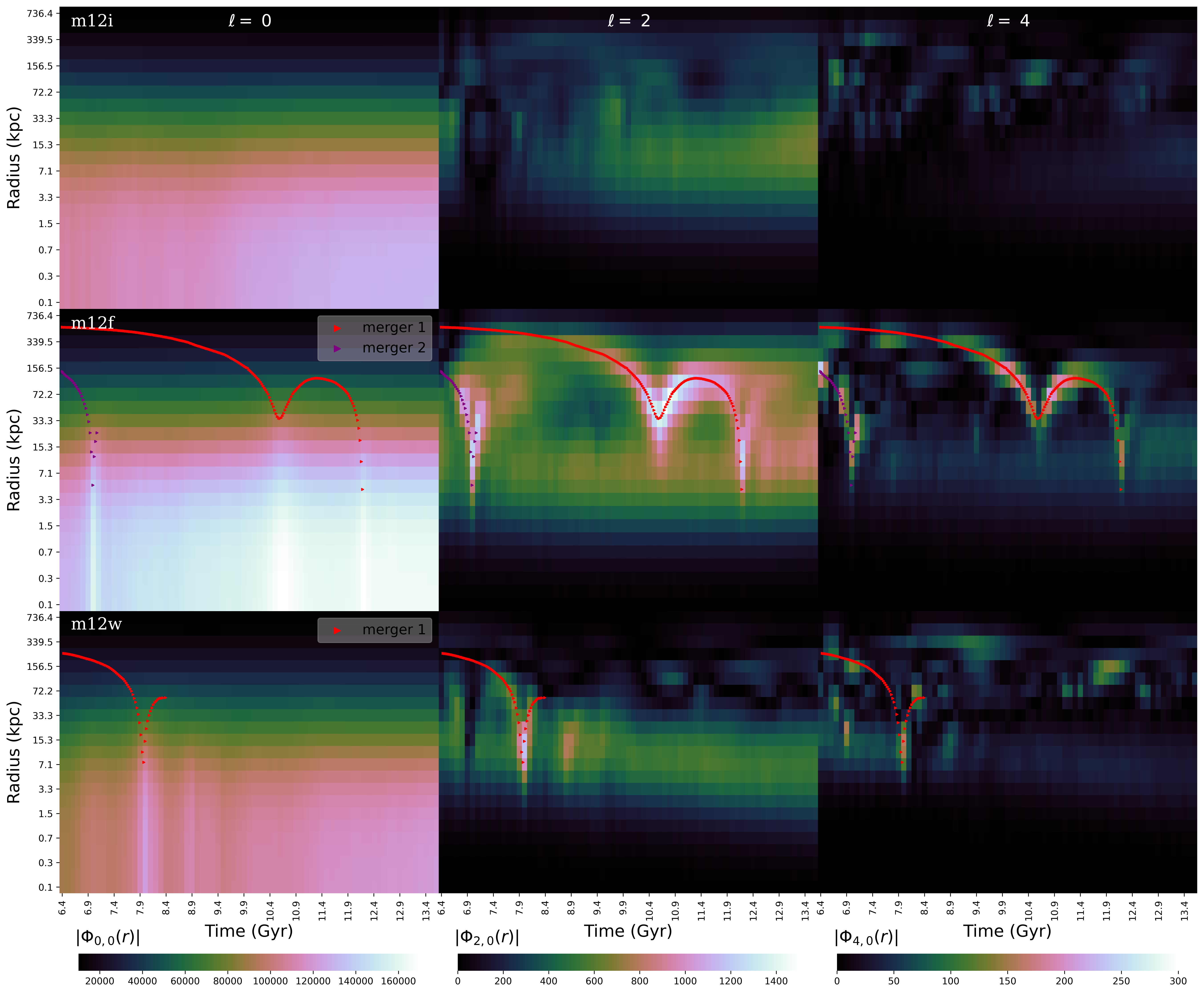}
    \caption{{Absolute value of {amplitude} $|\Phi_{\ell,m}|$ as defined in Eq.~\ref{eq:mult_exp} (color bar) at all radii as a function of time for the TEMP model for all the simulations. The TMR of these halos can be found in Table \ref{tab:merger}. The pericenter positions of the merging satellites are shown in red and purple in \mf{} (middle) and \mw{} (bottom). Due to axisymmetry, only the even poles exists with $m = 0$ components. Amplitude decrease as a function power law of radius with index $\sim -0.5$ for the zeroth pole which has all the information of the total mass of the halo. Higher poles have relatively low and uniformly distributed {amplitude} for \mi{} (top). In \mf{} and \mw{}, the pericenter position of the merging satellites (red and purple) strongly coincides with the radial locations with higher {amplitude} in the spectrum for the duration of the mergers for $\ell = 2,4$. This strongly suggests that even with axisymmetry constraints, the TEMP model with higher poles does well in modeling a dual COM system such as merging satellites.}}
    \label{fig:power_spec_all}
\end{figure*}

We start by examining the time-evolving multipole model for all the simulations. For each simulation, the expansion coefficients are calculated independently at each time snapshot at a range of galactocentric radii. Fig.~\ref{fig:power_spec_all} shows the absolute values of the expansion coefficients (color bar) $\Phi_{\ell,0}(r)$ (eq.~\ref{eq:mult_exp}) for even poles ($\ell \leq 4$) in the \mi{} (top), \mf{} (middle), and \mw{} (bottom) as a function of radius and time. For all the simulations, the $\ell = 0$ mode represents the spherically averaged mass profile of the galaxy, which is relatively stationary in time. The amplitude for $\ell = 0$ mode falls exponentially as a power law index of -0.4 with radius at the present day. On the other hand, higher poles show a wider variation between simulations, reflecting the differences in structure at the intermediate scales connected with LMC-like mergers. In \mi{}, which has primarily low-mass accretion at late times, the amplitude in the higher modes is evenly distributed in radius, with only some small perturbations. On the other hand, the merger information is clearly encoded in the higher poles for both the simulations with high-mass accretion (\mf{} and \mw{} ). We overplot the pericenter positions (red and purple) of the most massive merging subhalo[s] on Fig.~\ref{fig:power_spec_all}. For \mf{} and \mw, we see that the pericenter positions exactly coincide with the high-order perturbations present in the power spectrum (white regions). This behavior is seen in both \mf{} and \mw{} for $\ell \geq 2$ poles, suggesting the ability of the higher poles to accommodate the merger's effect on the overall mass distribution even though the model is constrained to be axisymmetric. 

Mergers with the lowest PMR (see Table \ref{tab:merger}) have the largest relative amount of {amplitude} in the $\ell \geq 2$ poles. As expected, the merger with the highest PMR, in \mi{}, shows no significant influence on the model. Conversely, the mergers in \mf{} and \mw, with much lower PMR, produce significant changes in the power spectrum (Fig.~\ref{fig:power_spec_all}). The {amplitude} in the higher poles of the potential model changes quickly in the region of the merger. While the merger in \mw{} is most rapid and leads to a complete disruption in a short time span of 0.3 Gyr, merger 1 in \mf{} takes about 2 Gyr, gradually varying the potential until it completely merges at 12.3 Gyr. Hence, this merger takes time to tidally strip stars and change the global potential profile of the Galactic halo. The timescale of the merger is longer compared to characteristic dynamical times at radii of interests, and to the timescale at which snapshots are saved, and thus the multipole expansion can more easily adapt at each time step to reflect the true potential.

This behavior is similar to the Gibbs phenomenon observed in Fourier series decomposition \citep{gibbs1898fourier}. The Gibbs phenomenon states that modeling a rapidly changing, discontinuous function using Fourier series (another orthogonal basis) leads  to large oscillations and overshooting around the jump discontinuity that cannot be removed by adding higher poles. The merger in \mw{} and merger 2 in \mf{} show this type of behavior: the Galactic potential profile has a rapid perturbation due to the merging halo. As the timescale of the merger approaches the time between snapshots, the perturbation approaches a discontinuity from the perspective of the potential model. This provides a fundamental limit to efforts to reproduce the evolving potential from a finite series of snapshots. On the other hand, the basis expansion has relatively little trouble when the merger is more slowly evolving, as in merger 1 in \mf{}, since the function remains relatively smooth throughout time and space.

\begin{figure}
    \includegraphics[width=\linewidth]{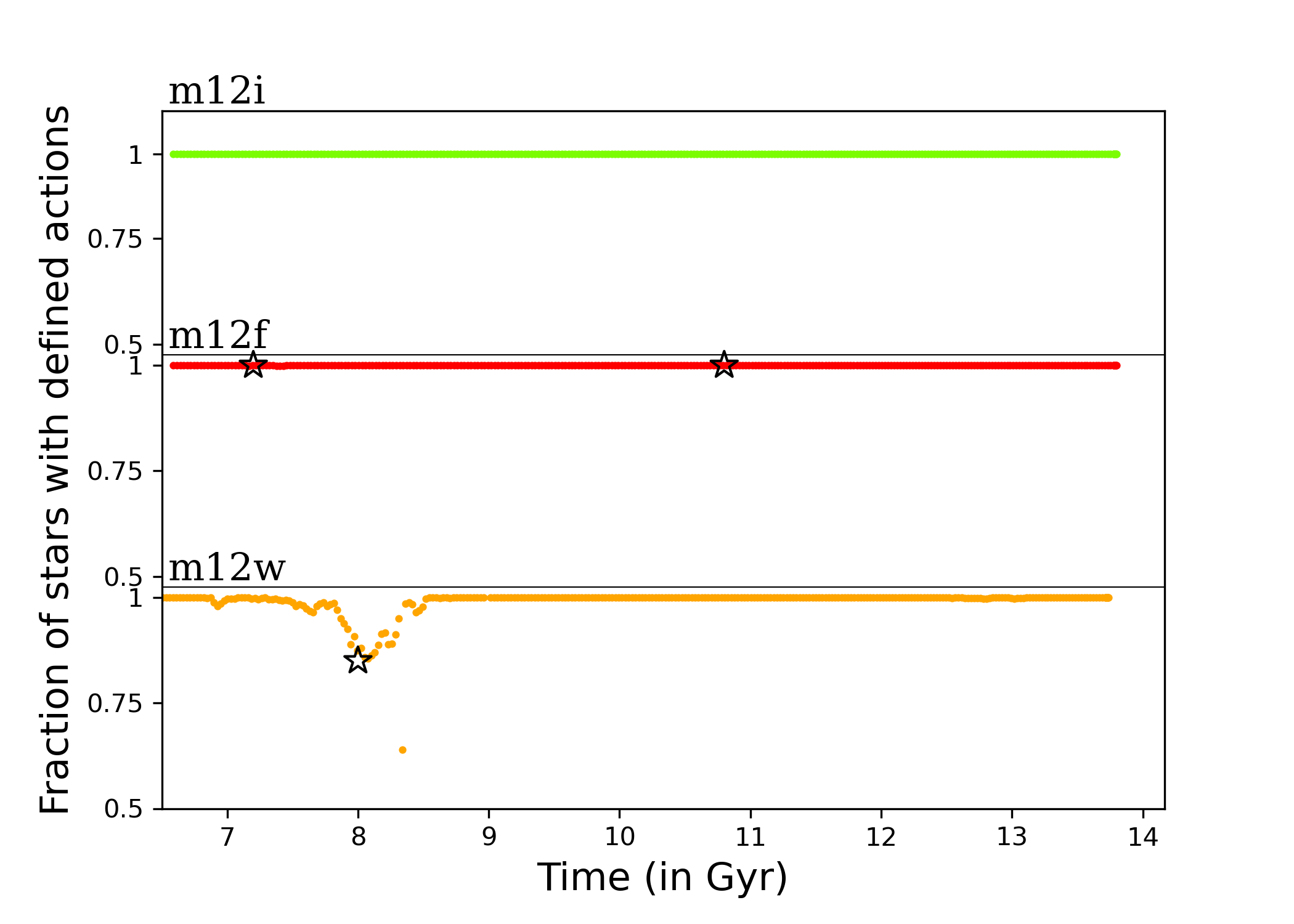}
    
    \caption{{Fraction of stars in the three simulations with defined actions, as a function of time in the TEMP model. The fraction is described by the total number of stars that have defined actions calculated using the TEMP model over the total number of stars bound to the main halo ($E \leq 0$). Note that for \mi{} (top) and \mf{} (middle), almost all of the stars have well-defined actions. While for \mw{} (bottom), about 20\% of stars do not have well-defined actions around the merger at 7.95 \textrm{Gyr}.}}
    \label{fig:frac_stars}
\end{figure}

\begin{figure}[!htb]
    \includegraphics[width=\linewidth]{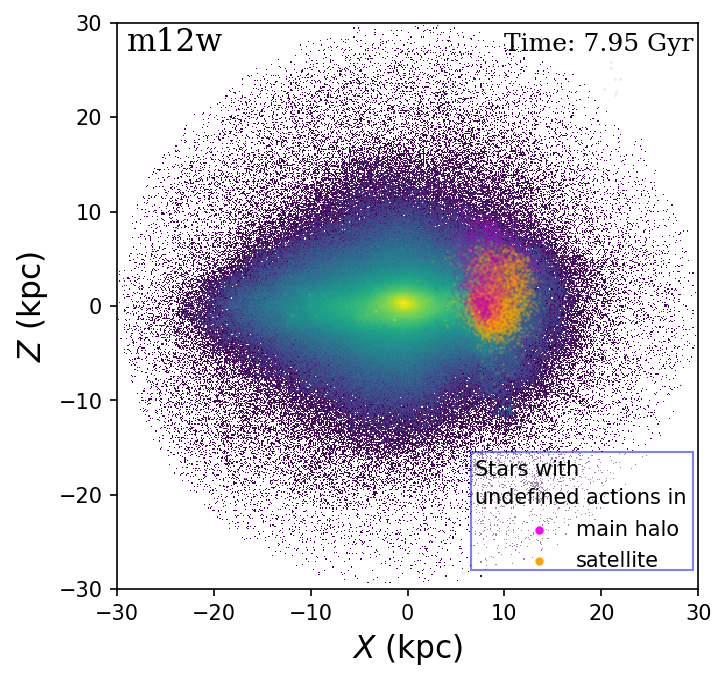}
    \caption{{Spatial distribution of stars in the \textit{Z-X} plane at 7.95 Gyr for \mw{}. Stars with undefined actions belonging to the main halo are shown in magenta and those belonging to the perturber are shown in yellow. All of these stars are in bound orbits around the main halo with $E \leq 0$ and in close proximity to the perturbing satellite.}}
    \label{fig:m12w_failed_actions}
\end{figure}

A useful global model for the MW potential would ideally be able to determine approximate actions for stars throughout the galaxy that reasonably describe their orbits. To quantify the first part of this requirement, in  Fig.~\ref{fig:frac_stars} we track the fraction of stars that are gravitationally bound to the main halo in each simulation (i.e. those for which actions should be defined) for which the axisymmetric BFE (specifically the TEMP model at $z=0$) can indeed determine defined actions. If the fraction drops below 1, it implies there are a substantial number of stars that the model potential incorrectly considers unbound, and hence cannot calculate their actions. In most cases, our BFE successfully calculates an action for every star, except in the merger in \mw{}, which has the lowest TMR and PMR. The fraction of stars with defined actions decreases around the merger time and falls to 0.6 at 8 \textrm{Gyr}, essentially implying a loss of nearly half of the the orbital information in the stars. We ascertain that this is due to the failure of the model to accommodate the merger by plotting the positions of the stars with undefined actions (magenta and yellow) in \mw{} along with the Galactic disk shown in Fig.~\ref{fig:m12w_failed_actions} at 7.95 Gyr (close to the time of the merger). The stars with failed actions are around the same radius as the perturbing satellite, which implies a failure to represent the potential localized near the perturber. These include both stars that are part of the merging satellite (yellow) and stars in the main halo at the same radius (magenta). This arises from the axisymmetry condition imposed on the potential model, as the perturber's mass is evenly distributed over a ring instead of being localized, which makes the kinetic energy of these particles to exceed the approximate potential energy. This effect is dominant for the massive satellite in \mw{} while the potential approximation with the symmetry condition works reasonably for \mi{} and \mf{} with less massive satellites.
We conclude that, at least as far as producing defined actions, a low-order axisymmetric BFE can successfully accommodate mergers with TMR $\gtrsim$ 10 and PMR $\gtrsim$ 3. In \S \ref{subsec:uniformKLD} and \S \ref{subsec:neighborKLD} we examine whether the actions usefully describe stellar orbits.

%##########################################################
%Figure : Power spectrum_all.
%#########################################################

While our BFE can usually accommodate a merger except in extreme cases, this is a bigger challenge for parameterized models. In the BFE the basis is chosen to automatically satisfy the Poisson equation, but for a parameterized model it is far more complex to add a second set of components to the potential with a moving center, since in order to be self-consistent the model must solve the Poisson equation at all times. The multipole expansion and the parameterized models both perform  well for simulations with no significant mergers, but when a relatively massive galaxy is near the host's center (TMR > 10 and/or PMR > 3) the multipole method is superior to parameterized models.  Even the TEMP models, which are highly effective in modeling Sgr-like mergers, do not perform as well for LMC-like mergers. 
%##########################################################
%Figure : Frac_stars.
%#########################################################

%##########################################################
%Figure : m12w_failed actions.
%#########################################################

\subsection{Action Space coherence}

The actions of stars in tidal streams should both be adiabatically invariant and remain significantly clustered for potential models that are good representations of the true potential of a halo. To test how well the BFE performs relative to a parameterized model, and whether it is important to model the time dependence in order to preserve the action space properties of streams, we use the various models we developed (Table \ref{tab:pot_models}) to calculate the actions of star particles in stellar streams in each simulated system at each snapshot from 7 Gyr ago to the present day. We then examine the clustering and stability of the action space distribution as a function of time.  

\begin{figure*}
    \includegraphics[width=\textwidth,]{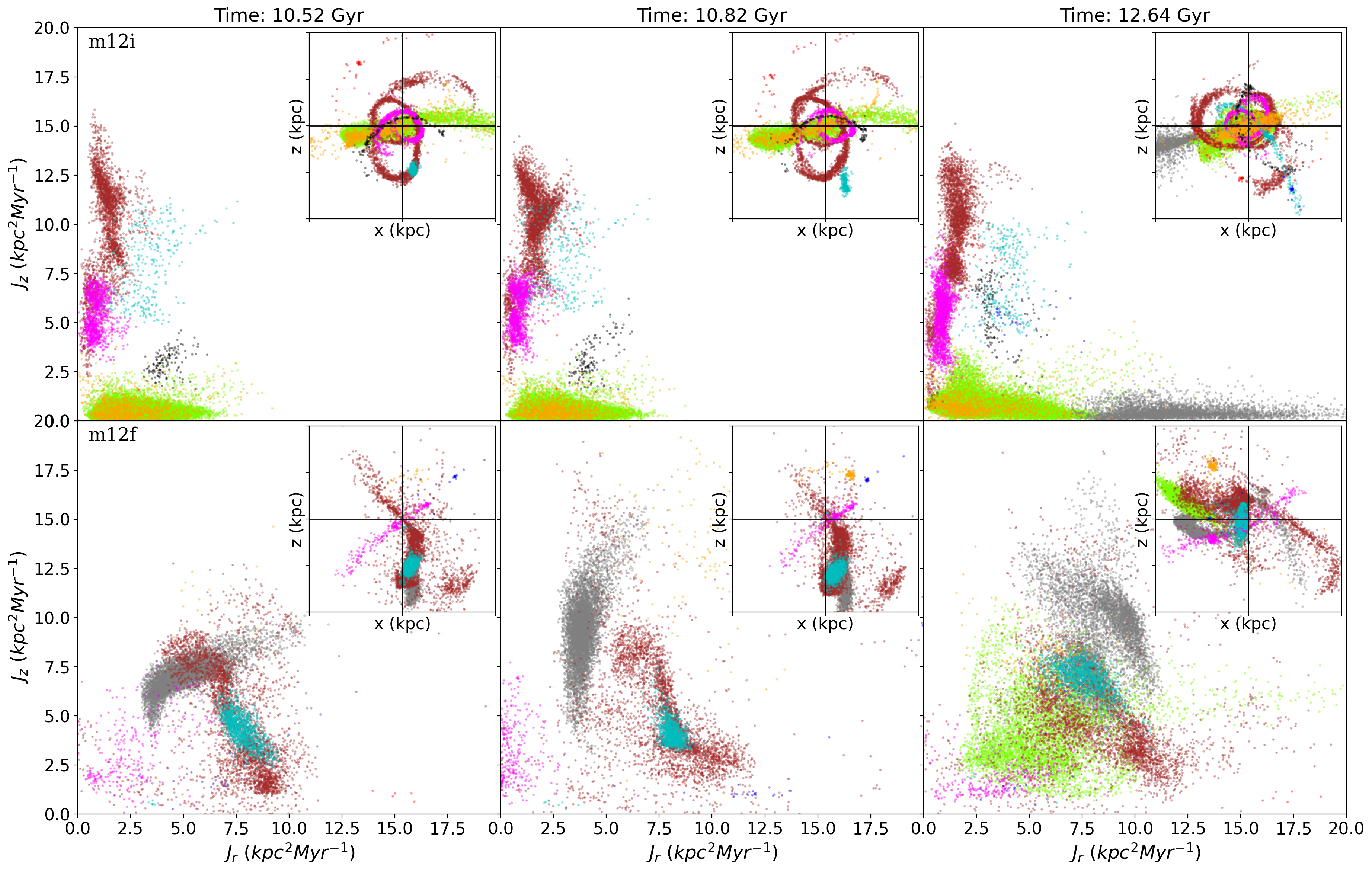}
    \caption{{Action space ($J_r$ vs. $J_z$) of stellar streams present in \mi{} (row 1) and \mf{} (row 2) found using the TEMP model. The three columns represent the time before the merger, during the merger, and after the merger in \mf{} at 10.52, 10.82 , and 12.64 Gyr, respectively. During the merger, the clustering of actions decreases in \mf{} (middle column). After the merger, the actions have been modified after the interaction (in \mf{}) and never retain their original values. While \mi{} has no such loss of clustering and instability in action space throughout different time frames. Note that a new stream forms in the third frame for both \mi{} and \mf{}. The complete time-evolving action space {{\href{https://drive.google.com/drive/folders/1t2dLAQx4X5H7WOrUhX9nyrf2_pdgD2ga?usp=sharing}{videos}}} are posted at \url{https://drive.google.com/drive/folders/1t2dLAQx4X5H7WOrUhX9nyrf2_pdgD2ga?usp=sharing} for all the models and simulations.}}
    \label{fig:actions_merger}
\end{figure*}

We visually analyze the action space by making {\href{https://drive.google.com/drive/folders/1t2dLAQx4X5H7WOrUhX9nyrf2_pdgD2ga?usp=sharing}{videos}} \citep{arora_dataset} of the action distribution for our simulations over the 7 Gyr period, using the actions computed in each of the potential models in Table \ref{tab:pot_models}. These {\href{https://drive.google.com/drive/folders/1t2dLAQx4X5H7WOrUhX9nyrf2_pdgD2ga?usp=sharing}{videos}} are posted at \url{https://drive.google.com/drive/folders/1t2dLAQx4X5H7WOrUhX9nyrf2_pdgD2ga?usp=sharing} and released as part of \cite{arora_dataset}. Fig.~ \ref{fig:actions_merger} shows three representative time frames (columns) of two independent actions ($J_r$ versus $J_z$) for \mi{} ($1^\textrm{st}$ row) and \mf{} ($2^\textrm{nd}$ row), computed using the TEMP model. The frames represent a time just before the PMR = 8 merger in \mf{}\ (10.52 Gyr, left column), during the merger (10.82 Gyr, middle column), and $>1$ dynamical time after the merger (12.64 Gyr, right column). Note that in both simulations, a new stream forms in the time between the middle and final frames shown. The action space distributions for both \mi{}\ and \mf{}\ at the end of the simulation (13.7 Gyr), about 1 Gyr (or roughly another $\sim 1-2$ dynamical times in the halo) after the final frame in this figure, are shown in Fig.~\ref{fig:samp_spacel}.

The actions for individual streams, marked by points in a single color for each stream, start out clustered in both galaxies, then visibly spread out in the case with the merger (\mf{}, bottom row) to a far greater extent than in the case without a merger (\mi{}, top row). The gray and red streams in \mf{} in particular show substantial decoherence in action space. The centroids of the individual action space clusters corresponding to the streams also move rapidly between snapshots in the {\href{https://drive.google.com/drive/folders/1t2dLAQx4X5H7WOrUhX9nyrf2_pdgD2ga?usp=sharing}{video}} of \mf{} during the merger. By the time the merger is finished (bottom right) the actions are permanently diffused and scrambled relative to their configuration prior to the merger. On the contrary, in \mi{} the actions of each stream's stars remain substantially more clustered around relatively stable centroids over the entire period of the {{\href{https://drive.google.com/drive/folders/1t2dLAQx4X5H7WOrUhX9nyrf2_pdgD2ga?usp=sharing}{(video)}}}. 

The decoherence of action space in \mf{} is clearly present in all four potential models, as expected given the low TMR and PMR.  On the other hand, \mi{} {\href{https://drive.google.com/drive/folders/1t2dLAQx4X5H7WOrUhX9nyrf2_pdgD2ga?usp=sharing}{videos}} show action space coherence at most times for all our models. Sample action spaces under different models for \mi{} and \mf{} are posted in Appendix.~\ref{app:action_models}. In Sec.~ \ref{subsec:uniformKLD} and \ref{subsec:neighborKLD}), we \emph{quantitatively} compare the degree of clustering and stability across potential models and simulations.   
% ########################################################
% FIGURE: Sample action space around merger
% ########################################################

%#########################################################
%                   KLD
%##########################################################

%##########################################################
%Uniform background KLD
%#########################################################

\subsection{Action space coherence of streams}
\label{subsec:uniformKLD}

We quantify the degree of clustering of the action space for various potential models by calculating the KLD relative to a uniform-background distribution, as discussed in section \ref{Sec:UKLD}, at each time step for all the potential models. A \emph{higher} KLD value represents a \emph{more clustered} action distribution. We refer to this quantity as the uniform-Background KLD or uKLD.

\begin{figure*}[!htb]
\includegraphics[width=\textwidth,]{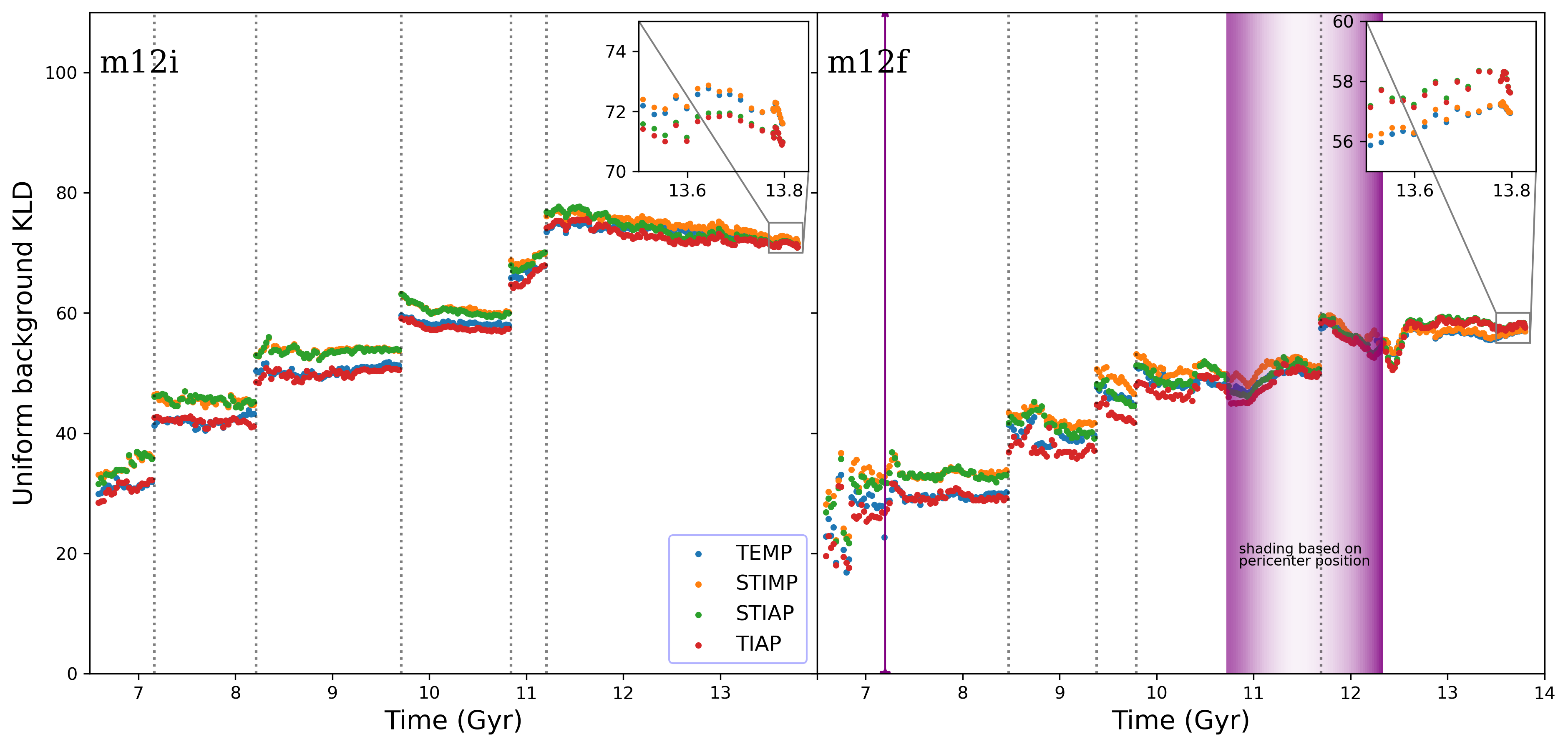}
\caption{{uKLD measures the action space clustering for streams in the simulations \mi{} (left) and \mf{} (right) from 6.8--13.8 \textrm{Gyr}. The two semi-TI models (orange and green) have systematically higher uKLD (a few nats) than the time-evolving model (blue) until all the stellar streams have formed. While the completely TI model (red) has the lowest uKLD. Discrete jumps in the uKLD are due to the formation of each new stream by tidal disruption; individual formation times are marked with gray vertical lines. While all the models produce a highly clustered action space around the present day, a drop in uKLD, indicating less action space clustering, can be clearly seen for \mf{} during both the mergers (purple line and purple shaded region). The prolonged merger in \mf{} is marked using a vertical purple band with the shading reflecting the proximity of the satellite to the Galactic Center (darker shades indicate smaller pericenter distance). The degree of clustering, and hence the accuracy of the potential model, decreases as the merging halo approaches the main halo. Fig.~\ref{fig:LMC_m12f} shows the exact pericenter distances for the two mergers in this simulation.}}
\label{fig:uni_kld}
\end{figure*}

Fig.~ \ref{fig:uni_kld} shows the uKLD for the four potential models, for \mi{} (left) and \mf{} (right). The orange points correspond with the STIMP (Section \ref{Sec:mult_model}), red and green points correspond with  TIAP and STIAP (Section \ref{Sec:anly_model}) models, respectively, while the blue points correspond to the TEMP model (Section \ref{Sec:mult_model}). The discrete jumps in the uKLD correspond to the addition of new stellar streams as they form; their tidal-disruption times as defined in \citet{panithanpaisal2021galaxy} are marked with a dark gray vertical line.
In the case of \mi{}, the uKLD for the TEMP model is systematically lower by about 3-4 nats compared to the TI models (except TIAP), until all the stellar streams are added. We note that this disparity arises from the mass ratio scaling at earlier time steps. We compute the actions ($J_r$ and $J_z$) of a particle of mass $m$, and (\textbf{x},\textbf{v}) at each time step using the model from the present day and then scaling them with the mass ratio at each time step. At earlier times, this ratio is lower than 1 which leads to an overall convergence in action values that appear as higher degree of clustering. As the halo stops growing and the mass ratio is close to 1, this disparity disappears after 11.5 \textrm{Gyr} and the KLDs become comparable with a difference of less than 1 nats between models. 

These differences are substantially less than the information contributed by a single stream, which is between 8 and 10 nats, and can be attributed to numerical integration of the uKLD. Thus we conclude that the action space has a high degree of clustering for all the models in \mi{}, which has no significant interactions during the time period. For such a scenario, a model with a fixed shape and adiabatically increasing mass appears sufficient to preserve the clustering; indeed, the STIMP model produces the highest clustering in the action space at all times for this simulation, although the TEMP model performs comparably to it once all of the streams are added.

%##########################################################
%Figure : uni back KLD
%#########################################################

The uKLD for \mf{} (Fig.~\ref{fig:uni_kld}, right panel) at first follows the same trend, where TEMP has a lower KLD value of about 3-4 nats until after a stream is added following the merger at 7.2 \textrm{Gyr}. The purple vertical lines and shaded band mark the times of mergers m12f-1 and m12f-2. Merger m12f-1 is long-lived and has a very small pericenter: the shading of the purple band corresponds to the distance of the merging galaxy from the host galaxy, with darker shades representing smaller distances. In all the models, the uKLD decreases substantially around the merger times. After the first dip in the uKLD at 7.2 \textrm{Gyr} (merger m12f-2), its value increases and stabilizes in all of our models. Subsequently, all the models have a relatively similar uKLD until the next merger (m12f-1), which lasts about 1.5 \textrm{Gyr} between 10.8 and 12.3 \textrm{Gyr}. A sudden dip in the uKLD is again apparent at the first pericenter of the merger, at 10.8 \textrm{Gyr} (darker shading). The uKLD then increases again as the merging satellite recedes from the host (lighter shading) and is highest when the merging halo is at the apocenter. Then as the satellite turns around, the uKLD decreases again more steeply than before, with a minimum at 12.2 \textrm{Gyr}. This second pericenter is closer than the first, and results in the complete tidal disruption of the satellite. Despite the difficulty in modeling the system when the merger is near pericenter, all models can produce a coherent action space for stellar streams with a similar degree of clustering, but time dependence and the flexibility of the potential model play a bigger role in this case than in the relatively stable \mi{} simulation: the STIAP (green) and TIAP (red) models lose the most information around the mergers (about 10 nats). Nonetheless, this example indicates that the BFE can indeed be used to model an SMC-mass merger, not only where the merging satellite is relatively far from the main halo, but even during and after a pericenter passage. Furthermore, although a  stream is added to the action space distribution during (indeed, as a result of) the merger, the uKLD at the end of the merger has decreased to roughly its pre-merger value again, essentially representing the loss of one stream's worth of information in action space as a result of the perturbation to the potential. This holds even after the merger is complete and the axisymmetric model should in principle be a good representation of the potential again. Note that example action spaces for all models are posted in Appendix \ref{app:action_models} and Fig.~\ref{fig:as_model_m12i} and \ref{fig:as_model_m12f} for \mi{} and \mf{} respectively, at two time steps: 13.8 Gyr and 8 Gyr. uKLD changes can be seen in the clustering of streams in this space.

\subsection{Orbital coherence}
\label{subsec:neighborKLD}

Even though our potential models produce action space clustering, the uKLD analysis does not capture the stability of the \emph{positions} of the streams in action space, which represents the degree to which the median or \emph{parent} orbit of the stream remains consistent. In a functional global model of the Galactic potential, the actions should also be {clustered under the assumption that the global potential profile is adiabatic}, and the stream clusters should therefore have stable positions in action space over cosmic time. The stability of the action distribution over time is thus a measure of the degree to which the quantities that we are calculating are true actions, in the sense that they reflect the stationary properties of the stellar orbits, or relatively arbitrary clustered quantities without a clear connection to orbits.            

\begin{figure*}[!htb]
\includegraphics[width=\textwidth,]{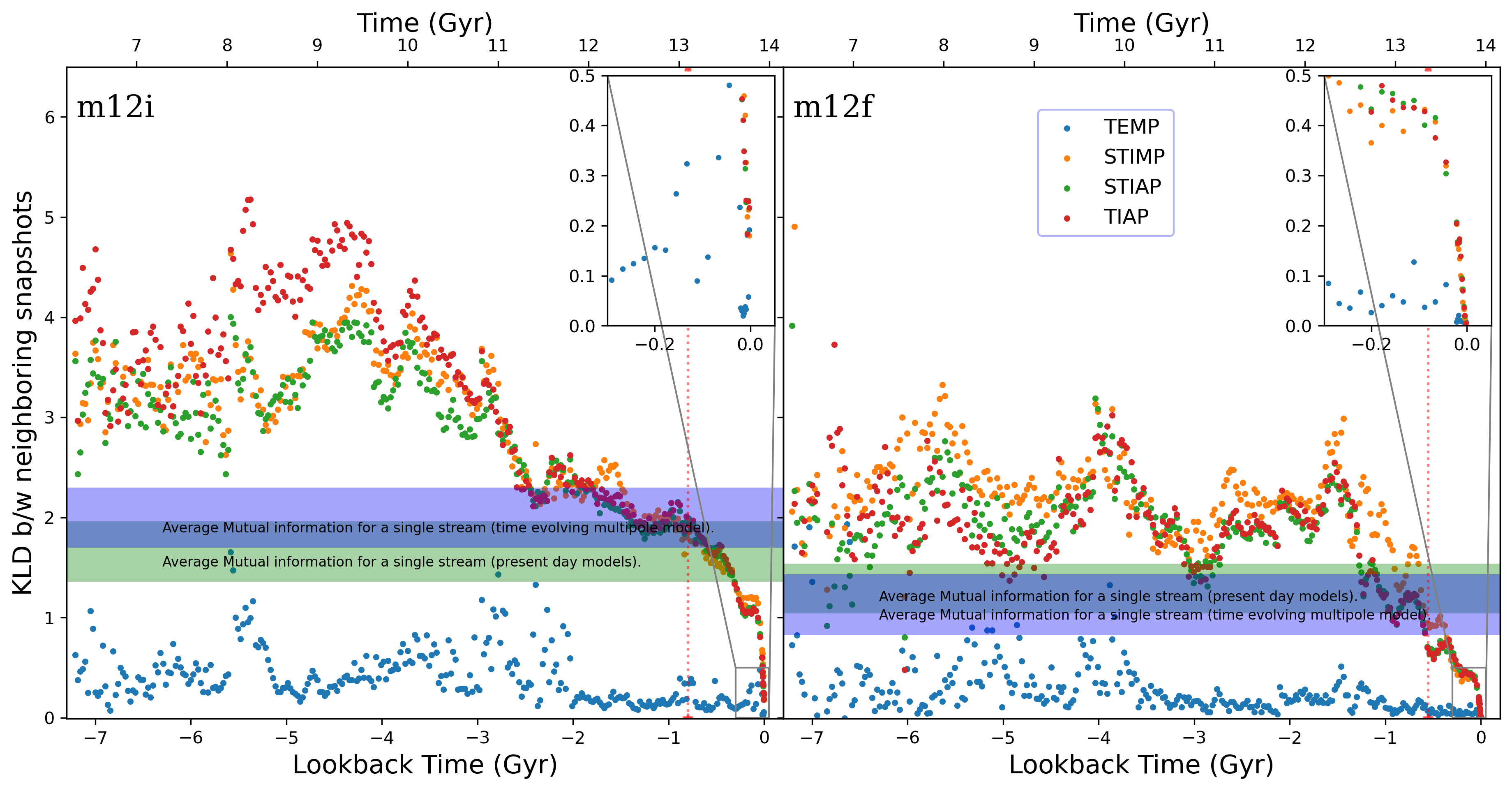}
\caption{{nKLD measures the stability of action space as a function of time for the simulations \mi{} (left) and \mf{} (right). For the TI models we compare the action space distribution at each snapshot to the present-day action space distribution, while for the TEMP model we compare the action space distributions of neighboring snapshots in time. For the TI models, the action space is completely decoherent (the nKLD is saturated) before 2.8 \textrm{Gyr}. For comparison, the shaded horizontal bands show the 95\%  range of MI computed for individual streams using the TEMP model (blue) and TI model (green), respectively. The time where our KLD value first exceeds the MI range, corresponding to an information loss equivalent to an entire stellar stream, is shown with a red dashed vertical line.}}
\label{fig:NN_kld}
\end{figure*}

To measure the stability of the action space distribution over time, we calculate the KLD between the action space distributions for pairs of neighboring snapshots in time for TEMP and KLD between each snapshot and present-day snapshot for TI models, as described in Section \ref{Sec:NKLD}, again using the actions computed for star particles in streams using each of our candidate potential models. We refer to this quantity as the neighboring KLD or nKLD. The snapshots are spaced by 25 Myr or less; i.e., much less than one dynamical time for the orbits in question. Thus, a low nKLD between snapshots corresponds to a small amount of nonadiabatic variation in the parent orbits of the streams, while a large nKLD indicates that the parent orbits are changing nonadiabatically. In Fig.~\ref{fig:NN_kld} we show the nKLD for all three potential models for \mi{} (left panel) and \mf{} (right panel). The higher nKLD values of the TI models indicate a lack of stability of its action space relative to the TEMP model (blue) in both simulations. We expect the nKLD of the TI models to increase with lookback time, as can be seen in Fig.~\ref{fig:NN_kld}, and this is true until about 2--3 Gyr in the past. Beyond this point, the nKLD oscillates around a roughly constant value, indicating complete decoherence of action space associated with a potential model fit to the present-day snapshot. In this regime, the action space of an individual snapshot calculated using a static model could still be clustered (as can be seen in Fig.~\ref{fig:uni_kld}), but the \emph{values} of the actions can no longer be interpreted as information about the parent orbit of a given stream at these earlier epochs. Conversely, the TEMP model preserves a relatively constant, low nKLD even beyond a 7 Gyr lookback time. Realistically, we cannot construct such a model from only present-day information about the MW, but here we can use it to quantify the information loss in TI models, to understand how far back in time we should expect a static model to produce actions that are connected to orbits.

For our TI models, the action space is completely decoherent before 2.8 \textrm{Gyr} and  1.5 \textrm{Gyr} ago for \mi{} and \mf{}, respectively. By this point, TI models have lost almost two-thirds of the information present, as the nKLD value increases from less than 1 to almost 3 for the TI models. Also as expected, the static potential models for \mi{} preserve a connection between actions and orbits to earlier lookback times than for \mf{}. For \mf{}, the nKLD value peaks at around 1.5 \textrm{Gyr}, the time at which the merger in that galaxy is complete, for all of our potential models. Orbits prior to the merger are not recoverable in the TI models, as the nKLD values oscillate randomly between 1 and 3. 

To interpret the nKLD as information loss, we compare it to the MI, defined in Section \ref{sec:MI}, in the action space distribution of each individual stream at the present day. This number quantifies the degree of correlation between actions that is expected for stars in a particular stream, and can be interpreted as the amount of \emph{information} contained in one stream. The horizontal bands in Fig.~\ref{fig:NN_kld} show the range containing 95\% of the MI values over all the streams in \mi{} and \mf{} for the TEMP (blue) and TI (green) models, respectively. As one might expect, the TEMP model preserves this correlation slightly better than the TI models do, and we see that each stream contains about 1.5--2.25 nats of information per star. When the nKLD approaches this number, it is as if about one stream's worth of information has been lost in the time between two snapshots \citep{sanderson2017modeling}. This occurs at somewhat smaller lookback times in both models than the time when the nKLD begins to saturate at complete decoherence: at about 0.8 and 0.5 \textrm{Gyr} ago for \mi{} and \mf{} respectively, for both TI models. Also as expected, the nKLD values for the TEMP models never exceed the MI of a single stream, with some exceptions very far in the past where few streams were yet formed.  

All our models perform relatively well around the present day. The insets in both panels of Fig.~\ref{fig:NN_kld} show the nKLD for the last $\sim20$ snapshots, including the final 10 up to the present day, which are spaced 2 Myr apart. Here, the TI models perform comparably to the TEMP model, although at larger lookback times the nKLD for the TEMP model is bounded while the nKLD for the two TI models diverges. The divergence near the present day is steeper in \mf{} because of the ongoing merger in that simulation. 

%##########################################################
%Figure : Next neighbor KLD
%#########################################################

We conclude that although the TI models can produce action spaces in which streams are clustered at relatively large lookback times, these actions cannot be connected with properties of orbits beyond a relatively short window into the past, since the clusters have shifted nonadiabatically from their present-day positions in action space. The TI models can represent the Galactic potential sufficiently well to preserve orbital properties furthest back into the past in cases with no large mergers, such as the \mi{} simulation. For these cases, the window in which TI modeling is sufficient appears to extend about 1.6 Gyr into the past. This window is shorter for more complicated potentials that include significant mergers: it reduces to less than 1 Gyr for \mf, and fails completely for \mw{}. 

Interestingly, given their single-origin expansion basis, the TEMP model demonstrates an exceptional ability to model potentials with a dual COM, as can be seen in Fig.~\ref{fig:power_spec_all}. Problems only arise where the PMR approaches {2}---{indicating that at pericenter the merger is behaving like a 2:1 merger}---or where the TMR is less than 10.  Even in those cases, the major issue is with the axisymmetry assumptions, which fail to find actions for stars where this assumption is extremely wrong. This usually occurs only for a short time right around pericenter and for a limited number of halo stars, as shown in Fig.~\ref{fig:frac_stars}. However, the BFE methods still work well when these symmetry assumptions are relaxed, at the cost of an increase in the number of coefficients.

\section{Conclusions}\label{Conc}

\label{Sec:Conc}

In this paper, we test different methods of galactic potential modeling by comparing how well they preserve stream orbital properties for a set of tidal streams created from disrupted satellite galaxies in zoomed cosmological-baryonic simulations. We compare an axisymmetric, time-evolving BFE potential model with more traditional static and/or parameterized models. Specifically, we use three TI potential models: an analytic potential model fit at the present day (TIAP), a TIAP model with a mass ratio scaling (STIAP) at all time steps, and a low-order ${\ell = 4}$ multipole model fit at the present day (STIMP) assuming axisymmetry with a mass ratio scaling; and a time-evolving low-order ${\ell = 4}$ multipole model (TEMP) where the expansion is computed at every time step in the simulation, also assuming axisymmetry. We also check the validity of our techniques for galaxies with a variety of merger histories and determine where even the best potential method breaks down.

We find that the multipole expansion can accurately model potential varying mergers with satellites such as the LMC and Sgr dSph by using the higher pole orders of the expansion, even when restricted to axisymmetry. The low-order axisymmetric expansion preserves orbit structure and action-space coherence for stars in stellar streams for satellites that have the total merger ratio (the ratio between the total mass of the main halo and merging halo at the first pericenter) greater than 10, or a pericenter merger ratio (the ratio between the mass of the main halo enclosed within the pericenter position of the merging halo to the mass of the merging halo at the first pericenter) greater than 1 and have properties such as low radial velocities (order of 10 km/s) and pericenters outside the scale radius of DM halo. A good observational example of such a merger is the Sgr dSph which is thought to have a TMR greater than 10 \citep{niederste2010re}. When the model does break down and lose significant orbital information, it is mainly due to a failure in the axisymmetry assumption and could be evaded by relaxing the symmetry, which however leads to added difficulties in computing actions. Interactions such as the ongoing merger of the LMC and MW, which has an estimated TMR of 7.6 \citep{penarrubia2015_LMC,vasiliev2021tango}, are thus not expected to be well modeled by an axisymmetric low-order multipole with a single expansion center.

All types of models we tested could preserve the clustering of stream stars in action space. In cases with a large merger (by the above criteria) the degree of clustering decreases only temporarily during the merger. However, only the TEMP model demonstrated long-term orbital stability, defined as the consistency of the positions of the action-space clusters. All TI models demonstrate a total loss of information as early as 1 Gyr in the past for the simulations with no prominent mergers and about 0.7 Gyr in the past in the presence of even a Sgr-like merger, less than the dynamical times for the orbits of stream stars $\sim 1-3$ Gyr. Orbits for objects in the MW halo projected outside this time window are likely to diverge significantly from reality.

\newpage
\section*{Acknowledgements}{
The authors thank Denis Erkal and Eugene Vasiliev for the valuable discussions that shaped this paper. {We would also like to acknowledge the anonymous referee for their constructive feedback and input which helped us interpret and present our results better.}  

A.A. and R.E.S. acknowledge support from the Research Corporation through the Scialog Fellows program on Time Domain Astronomy, from NSF grant AST-2007232, and from NASA grant 19-ATP19-0068. R.E.S. additionally acknowledges support from  HST-AR-15809 from the Space Telescope Science Institute (STScI), which is operated by AURA, Inc., under NASA contract NAS5-26555.
AW received support from NSF via CAREER award AST-2045928 and grant AST-2107772; NASA ATP grant 80NSSC20K0513; HST grants AR-15809, GO-15902, GO-16273 from STScI.

This research is part of the Frontera computing project at the Texas Advanced Computing Center (TACC). Frontera is made possible by the National Science Foundation award OAC-1818253.  Simulations in this project were run using Early Science Allocation 1923870, and analyzed using computing resources supported by the Scientific Computing Core at the Flatiron Institute.  This work used additional computational resources of the University of Texas at Austin and TACC, the NASA Advanced Supercomputing (NAS) Division and the NASA Center for Climate Simulation (NCCS), and the Extreme Science and Engineering Discovery Environment (XSEDE), which is supported by National Science Foundation grant number OCI-1053575.

This work was performed in part at the Aspen Center for Physics, which is supported by National Science Foundation grant PHY-1607611. 

This research was initiated at the Kavli Institute for Theoretical Physics workshop “Dynamical Models for Stars and Gas in Galaxies in the Gaia Era”, supported in part by the National Science Foundation under grant No. NSF PHY-1748958.

FIRE-2 simulations are publicly available \citep{wetzel2022public} at \url{http://flathub.flatironinstitute.org/fire}. Additional FIRE simulation data is available at \url{https://fire.northwestern.edu/data}. A public version of the \textsc{Gizmo} code is available at \url{http://www.tapir.caltech.edu/~phopkins/Site/GIZMO.html}.

}

\vskip0.2in
\bibliography{ref}
\bibliographystyle{aasjournal}

\appendix

In this appendix, we present additional figures for reader's reference. Appendix  ~\ref{app:acc_m12i} compares the true central accelerations with spline accelerations for the three Cartesian axes and Appendix ~\ref{app:action_models} compares the action spaces for all the potential models at two different time steps: (i) (13.8 Gyr) the present day, (ii) (8 Gyr) a time far before present day for \mi{} and \mf{}.

\subsection{Spline fits for COM acceleration}\label{app:acc_m12i}

\begin{figure*}[!h]
    \includegraphics[width=\textwidth,height=\textheight,keepaspectratio]{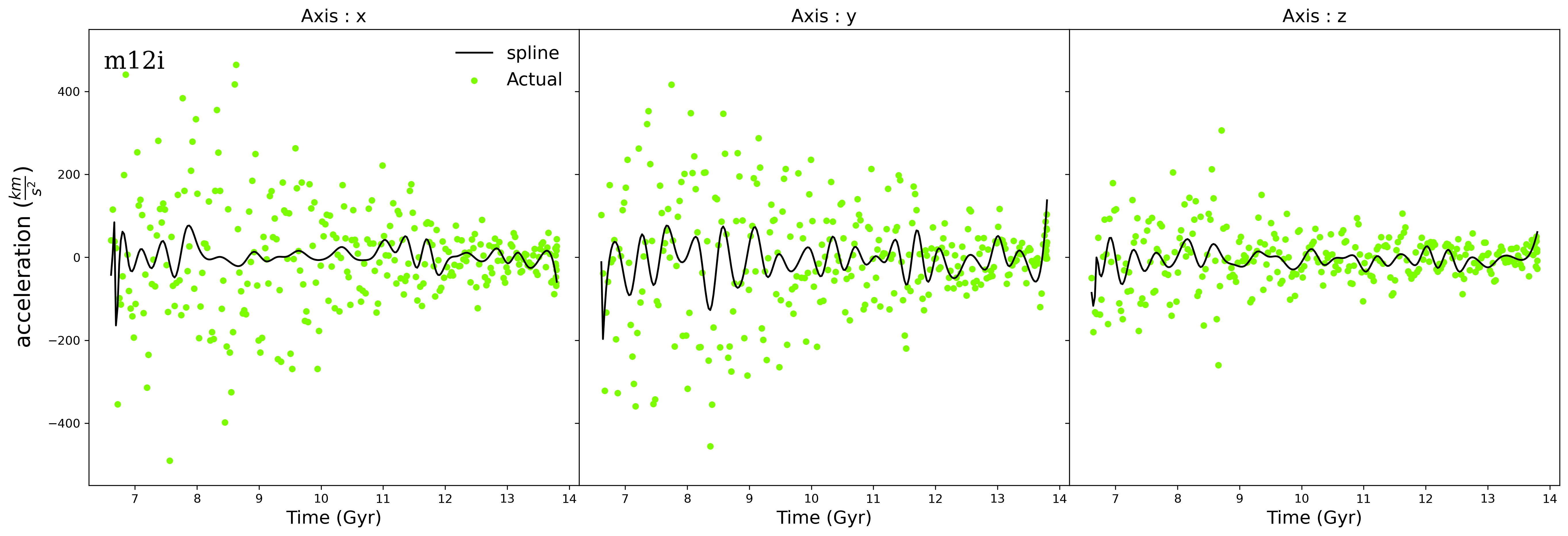}
    \caption{The three Cartesian accelerations (green) of the COM position along with the spline accelerations (black) calculated using the smooth cubic spline fits for each coordinate axis for \mi{}. The actual acceleration is jittery, highly oscillatory, and large while the spline values are significantly smooth and lower in comparison. The spline fit COM positions do not adversely affect the action space with fictitious forces arising due to jittery COM and only depend on the potential model.}
    \label{fig:acc_app}
\end{figure*}

Fig.~~\ref{fig:acc_app} plots the central accelerations (green) computed using numerical differentiation along with the spline acceleration (black) (refer to section~\ref{sec:centering}), computed from the spline fits on the COM position along each Cartesian axes for \mi{}. Note that actual central accelerations are jittery, oscillatory, and large while the spline accelerations are relatively lower. Therefore, we use the spline centering technique to fit our potential models, which makes the overall system less non-conserved in comparison with the actual accelerations. These jitters in actual accelerations are caused due to active star formation near the center \citep{orr2021fiery}.

%##########################################################
%Figure : Acc_example m12i.
%#########################################################

\subsection{Evolution of Action Space for All models} \label{app:action_models}

\begin{figure}[H]
    \includegraphics[width=\textwidth,]{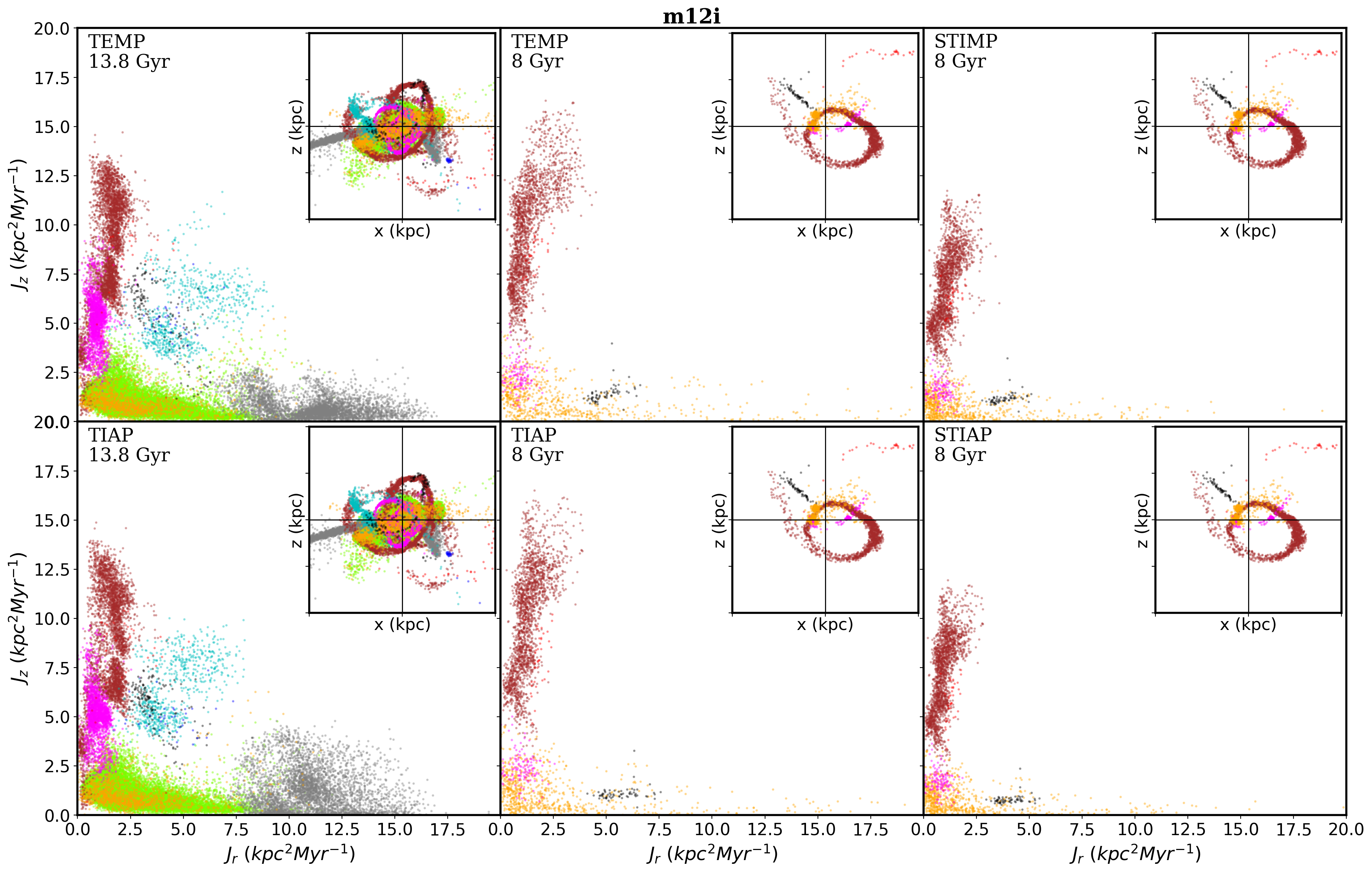}
    \caption{Action-space ($J_r$ vs. $J_z$) of stellar streams present in \mi{} for all the models (labeled on the top left) at two time steps, 13.8 and 8 Gyr respectively. The complete time-evolving action-space {\href{https://drive.google.com/drive/folders/1t2dLAQx4X5H7WOrUhX9nyrf2_pdgD2ga?usp=sharing}{videos}} are posted for all the models and simulations. The actions stay relatively consistent and clustered at both time steps for all the models. Although the semi-TI model produce higher clustering at 8 Gyr, which is due to the (mass ratio < 1) scaling at earlier time steps.}
    \label{fig:as_model_m12i}
\end{figure}

\begin{figure}[H]
    \includegraphics[width=\textwidth,]{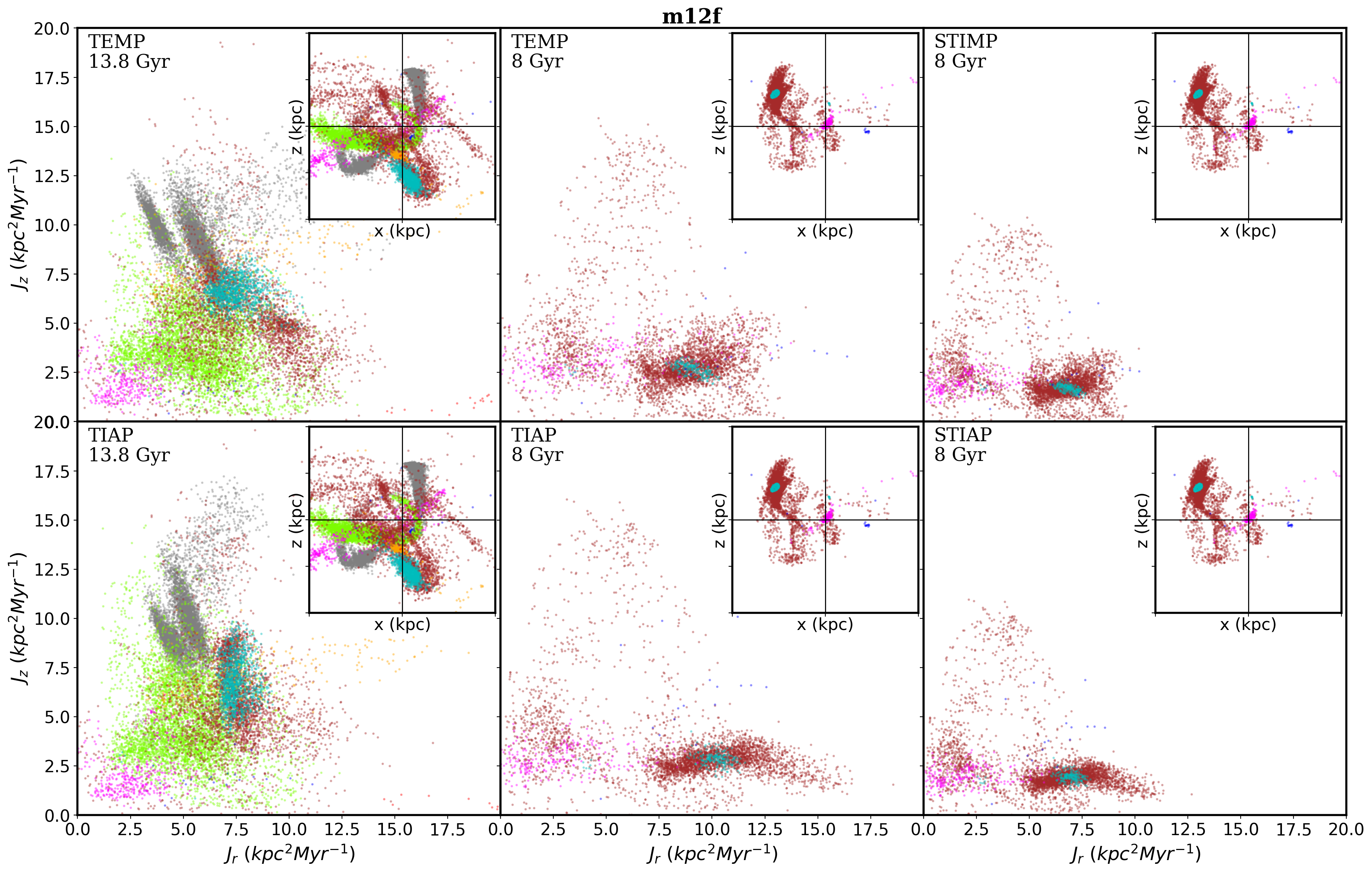}
    \caption{Action space ($J_r$ vs. $J_z$) of stellar streams present in \mf{} for all the models (labeled on the top left) at two time steps, 13.8 and 8 Gyr respectively. The complete time-evolving action-space {\href{https://drive.google.com/drive/folders/1t2dLAQx4X5H7WOrUhX9nyrf2_pdgD2ga?usp=sharing}{videos}} are posted for all the models and simulations. The action clusters are completely incoherent (lose clustering and change positions) at 13.8 Gyr compared to 8 Gyr due to a potential varying merger at 10.8 Gyr.}
    \label{fig:as_model_m12f}
\end{figure}

Fig.~\ref{fig:as_model_m12i} and \ref{fig:as_model_m12f} plot the action space ($J_r$ vs $J_z$) of stellar streams at 13.8 Gyr (Column 1) and 8 Gyr (Column 2 and 3) for the two multipole potential models described in Sec.~ \ref{Sec:mult_model} (row 1) and the two analytic models described in Sec.~ \ref{Sec:anly_model} (row 2) for \mi{} and \mf{} respectively. The insets on the top right plot the position-space distribution in the \textit{X--Z} plane. All the models produce highly clustered actions at the present day; however, multipole models can be seen to have higher clustering at 13.8 Gyr for \mi{}(red and gray streams) while it is hard to visually comment on the degree of clustering incase of \mf{}. At 8 Gyr, semi-TI models produce more clustering, which is an artifact of scaling the actions with the mass. The mass ratio at earlier snapshots is much less than 1, and multiplying this scaling with actions increases the degree of clustering; as it scales downs the action values. Note the difference between TIAP and STIAP models at 8 Gyr as shown in Fig.~\ref{fig:as_model_m12i} and ~\ref{fig:as_model_m12f}. For e.g. the STIAP action space is computed by scaling the TIAP action space with the mass ratio which inherently increases the clustering. The relative positions of action clusters change significantly for \mf{} from 8 and 13.8 Gyr due to a potential varying satellite merger, while the positions remain consistent for isolated galaxies with no mergers in the case of \mi{}. 

% ########################################################
% FIGURE: action-space all models (m12i)
% ########################################################

% ########################################################
% FIGURE: action-space all models (m12f)
% ########################################################

\end{document}